\documentclass[12pt,preprint]{aastex}

\usepackage{graphicx, url}

\slugcomment{Accepted by MNRAS}

\shorttitle{Head-Tail Clouds}
\shortauthors{Putman et al.}

\def\gtrapprox{\;\lower 0.5ex\hbox{$\buildrel >\over \sim\ $}}
\def\lessapprox{\;\lower 0.5ex\hbox{$\buildrel < \over \sim\ $}}

\def\Msun  {${\rm M}_\odot$}
\def\deg   {$^\circ$}

\def\kms   {\ km s$^{-1}$}
\def\Mhi   {M$_{\rm HI}$}

\begin{document}

\title{Head-Tail Clouds:  Drops to Probe the Diffuse Galactic Halo}

\author{M. E. Putman, D. R. Saul \& E. Mets} 
\affil{Department of Astronomy, Columbia University, New York, NY 10027; mputman@astro.columbia.edu}

\begin{abstract}
A head-tail high-velocity cloud (HVC) is a neutral hydrogen halo cloud that appears to be interacting with the diffuse halo medium as evident by its compressed head trailed by a relatively diffuse tail. 
This paper presents a sample of 116 head-tail HVCs across the southern sky ($d < 2$\deg) from the HI Parkes All Sky Survey (HIPASS) HVC catalog, which has a spatial resolution of 15.5\arcmin~(45 pc at 10 kpc) and a sensitivity of N$_{\rm HI}=2 \times 10^{18}$ cm$^{-2}$ ($5\sigma$).  35\% of the HIPASS compact and semi-compact HVCs (CHVCs and :HVCs) can be classified as head-tail clouds from their morphology.  The clouds have typical masses of 730 \Msun~at 10 kpc (26,000 \Msun~at 60 kpc) and the majority can be associated with larger HVC complexes given their spatial and kinematic proximity.   This proximity, together with their similar properties to CHVCs and :HVCs without head-tail structure, indicate the head-tail clouds have short lifetimes, consistent with simulation predictions. 
Approximately half of the head-tail clouds can be associated with the Magellanic System, with the majority in the region of the Leading Arm with position angles pointing in the general direction of the movement of the Magellanic System.   The abundance in the Leading Arm region is consistent with this feature being closer to the Galactic disk than the Magellanic Stream and moving through a denser halo medium.  The head-tail clouds will feed the multi-phase halo medium rather than the Galactic disk directly and provide additional evidence for a diffuse Galactic halo medium extending to at least the distance of the Magellanic Clouds.  
\end{abstract}


\section{Introduction}

A Milky Way-sized galaxy is expected to have a hot ($\sim10^6$ K), diffuse, halo medium that extends out to $\sim150$ kpc 
according to models of galaxy formation \citep{white78, maller04, fukugita06, kaufmann06, sommer06}.  Observationally this gas is extremely difficult to detect directly and has largely been inferred to exist from indirect methods, including: high velocity O~VI absorption as halo clouds interact with this medium \citep{sembach03}, the detection of diffuse x-ray emission \citep{wang93, rasmussen09}, O~VII and O~VIII absorption line results \citep{williams05,wang05}, and
dispersion measures of pulsars at a range of distances \citep{taylor93, gaensler08}.   Another important probe of the diffuse Galactic halo is the structure of the HI halo clouds, or high velocity clouds (HVCs), that indicate the clouds
are being both supported and destroyed by the surrounding halo medium \citep{peek07, putman03, stanimirovic02, wolfire95}.  In particular, the head-tail HVCs show a compact head trailed by a generally thinner and more diffuse tail \citep{bruens00, westmeier05b} and these structures are naturally produced in simulations of clouds moving through a lower density halo medium (Heitsch \& Putman 2009 (hereafter HP09); Quilis \& Moore 2001; Konz et al. 2002)\nocite{heitsch09, quilis01, konz02}.    Since evidence for the diffuse halo medium is observed in the clouds associated with the Magellanic System \citep{sembach03,putman03,stanimirovic02} and Milky Way satellites are stripped of their gas out to large radii \citep{grcevich09, blitz00}, the hot halo medium extends out to at least 60 kpc and may be unlikely to have all risen from the Galactic disk.

The origin of the HVCs found moving through this diffuse halo medium remains another unknown that can be addressed through further studies 
of their structure and properties.   The most likely origins for the clouds are debris from satellites and/or the cooling of over-dense regions in the multiphase halo medium.  The structure of the head-tail clouds can provide information on the disruption of the clouds, their lifetimes, and the direction they are moving.   If a large number of clouds are being disrupted on a relatively rapid timescale, a continuous formation mechanism may be needed to account for the numerous HVCs we observe today.    If the clouds show evidence for moving in the direction of the Galactic Plane, this suggests net infall of the population, and if they are pointing away from the Plane, this would indicate ejection from the disk.  Other net directions may indicate an initial motion for the clouds was set by a progenitor, such as a satellite.  The relationship of the head-tail clouds to other high velocity complexes also provides insight into their origin.  If the majority of the clouds are in the vicinity of other complexes, they may largely represent sheared off structures, while isolated head-tail clouds could represent the final remnants of a larger complex, or a more recently formed cloud that is just beginning to fall through the halo.
We also might expect the properties of the clouds to vary according to their densities and velocities and the density of the surrounding halo medium (HP09; \cite{gunn72}).  

In this paper we present the properties of the head-tail HVCs found in the HI Parkes All Sky Survey (HIPASS) catalog of southern clouds (\cite{putman02}; hereafter P02).  This paper is complemented by a paper that is capable of reproducing these clouds using three-dimensional grid simulations (HP09). The only other observational study of a substantial population of head-tail clouds was of the northern sky and was limited in its ability to identify cloud features due to the 
36\arcmin\ resolution (Br\"uns et al.~2000). 
The selection of the head-tail clouds from the HIPASS HVC catalog is described in Section 2,  the properties of the sample in Section 3, and a discussion of the results in the context of their origin and impact is presented in Section 4.

\section{Head-tail Cloud Selection}

The head-tail clouds were selected from the HVC catalog created from HIPASS data (P02).  The HIPASS HVC data were reduced with a modified version
of the standard HIPASS reduction software in order to recover extended emission,
referred to as the {\sc minmed5} method (see \cite{putman03} for details).  The HIPASS
HVC catalog covers the entire RA range, declinations from $-90$\deg~to $+2$\deg, and velocities relative to the Local Standard of Rest between $500 > |$V$_{\rm LSR}| \gtrapprox 90$ \kms.  Since the $\pm$90 \kms~cut in LSR velocity did not exclude all gas that is clearly part of the Milky Way at low Galactic latitudes,
an additional cut based on
the deviation velocity, or the amount the velocity of a cloud deviates
from a simple model of Galactic rotation, of $|$V$_{\rm dev}| > 60$ \kms~was
applied.
 The {\sc minmed5} reduced HIPASS data has a spatial resolution of
15.5\arcmin\ and a spectral resolution, after Hanning smoothing, 
of 26.4 km s$^{-1}$.   For extended sources, the RMS noise is 10 mJy
beam$^{-1}$ (beam area 243 arcmin$^2$), corresponding to a brightness
temperature sensitivity of 8 mK and a column density sensitivity of
$4 \times 10^{17}$ ($1\sigma$) per 26 \kms~resolution element. 
At 10 kpc the data has a sensitivity to 
clouds of gas with \Mhi\ $\sim 100$ \Msun\ ($\Delta$v = 26.4 \kms; 5$\sigma$) and the
beam is 45 pc, and at 60 kpc this mass limit is \Mhi\ $\sim 3500$ \Msun\ and the beam is 270 pc.  As often found for catalogs, the P02 cloud catalog may not be complete until values 5-6 times these values.

\begin{figure}[h!]
\begin{center}
\vspace{-0.4in}
\includegraphics[height=2.5in]{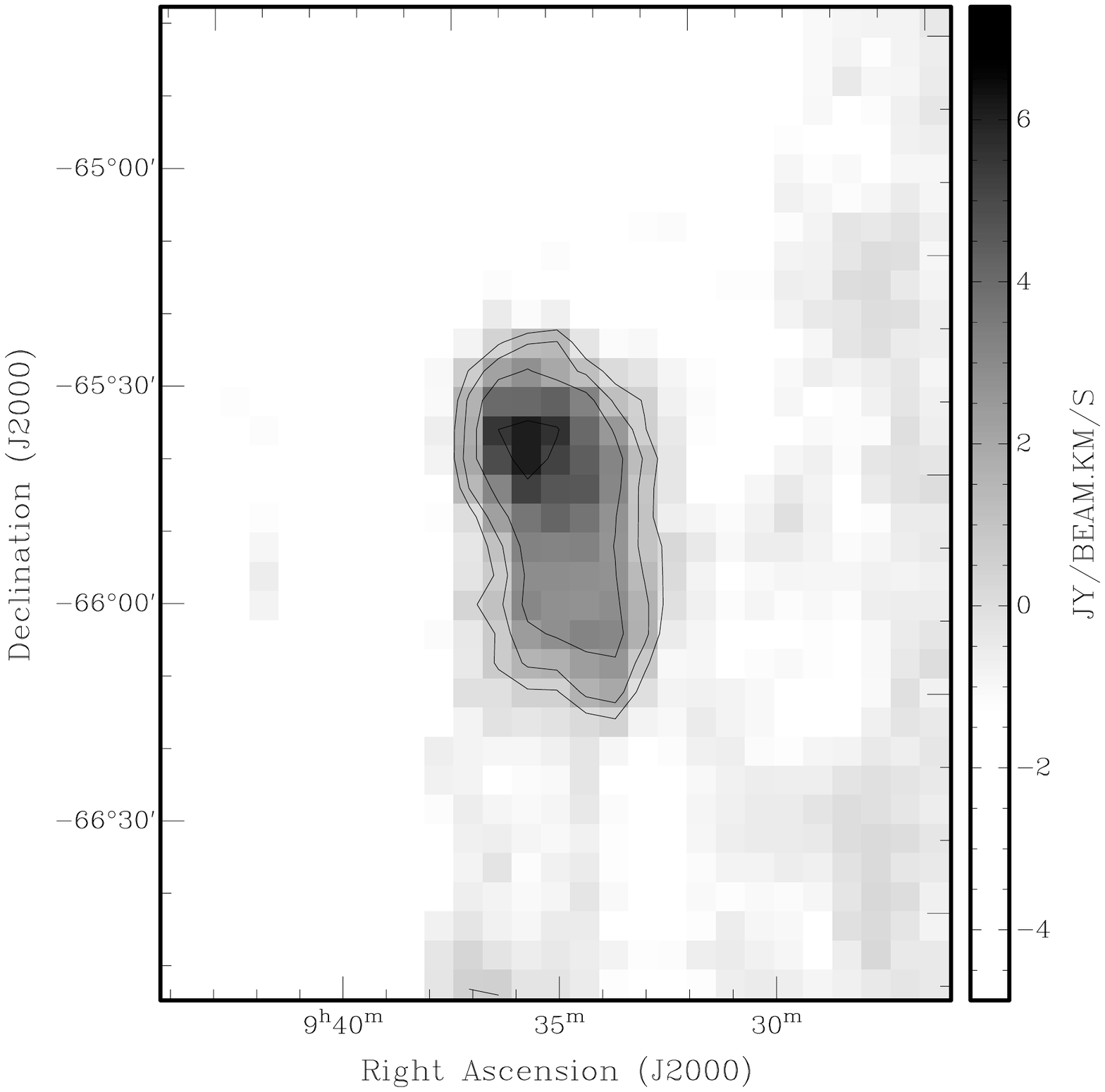}
\vspace{-0.8in}
\includegraphics[height=2.5in]{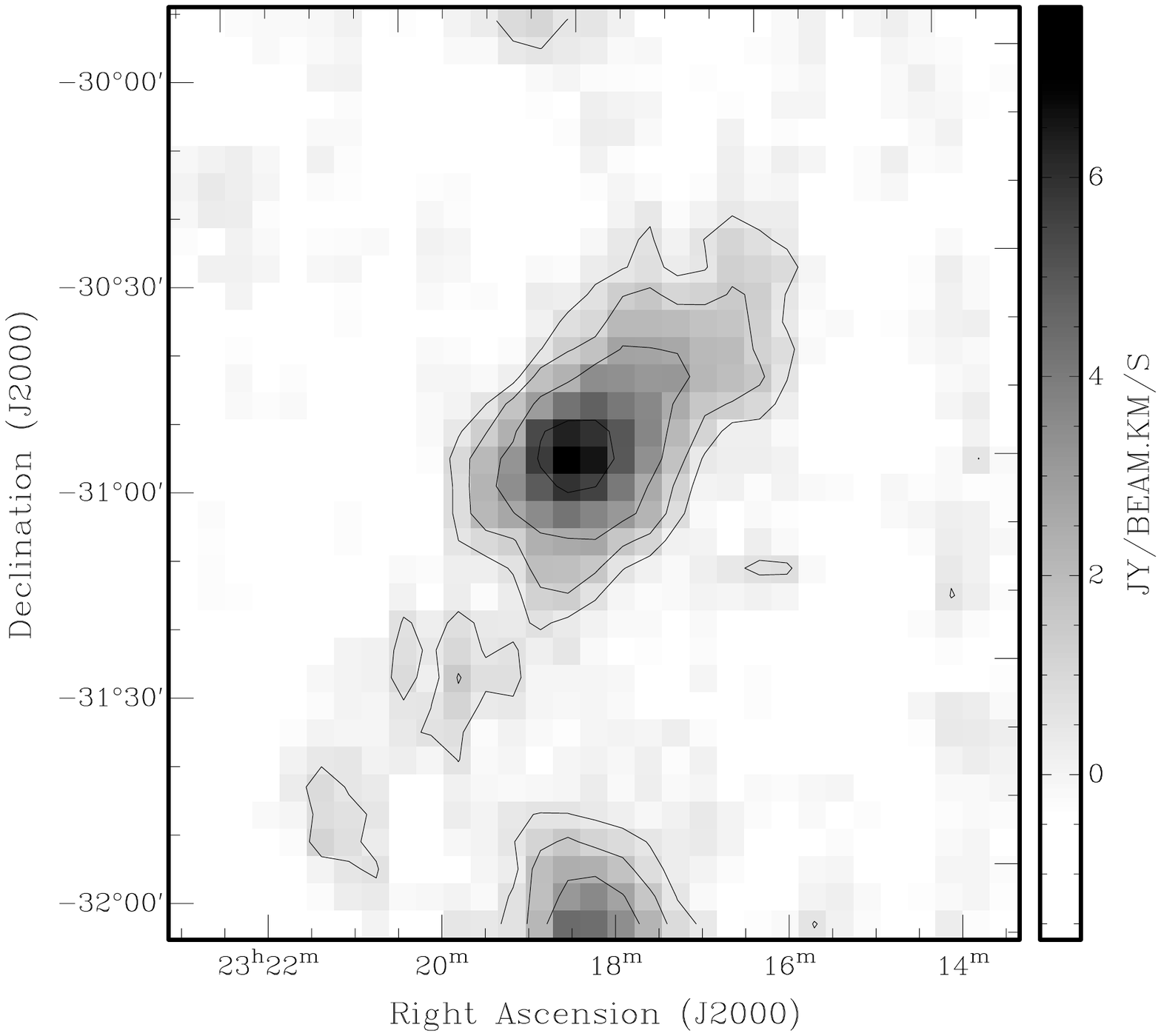}
\includegraphics[height=2.5in]{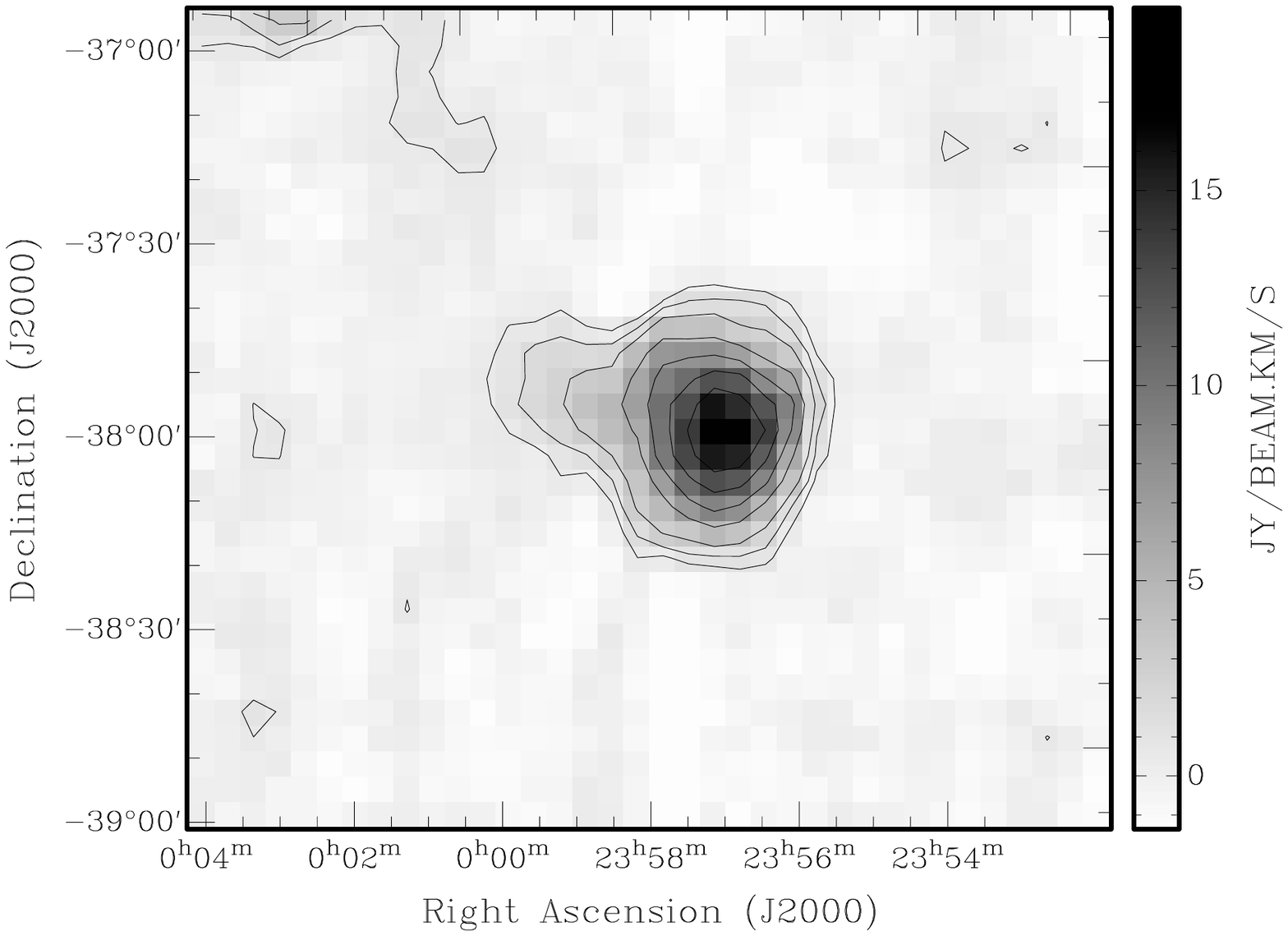}
\vspace{-1in}
\includegraphics[height=2.5in]{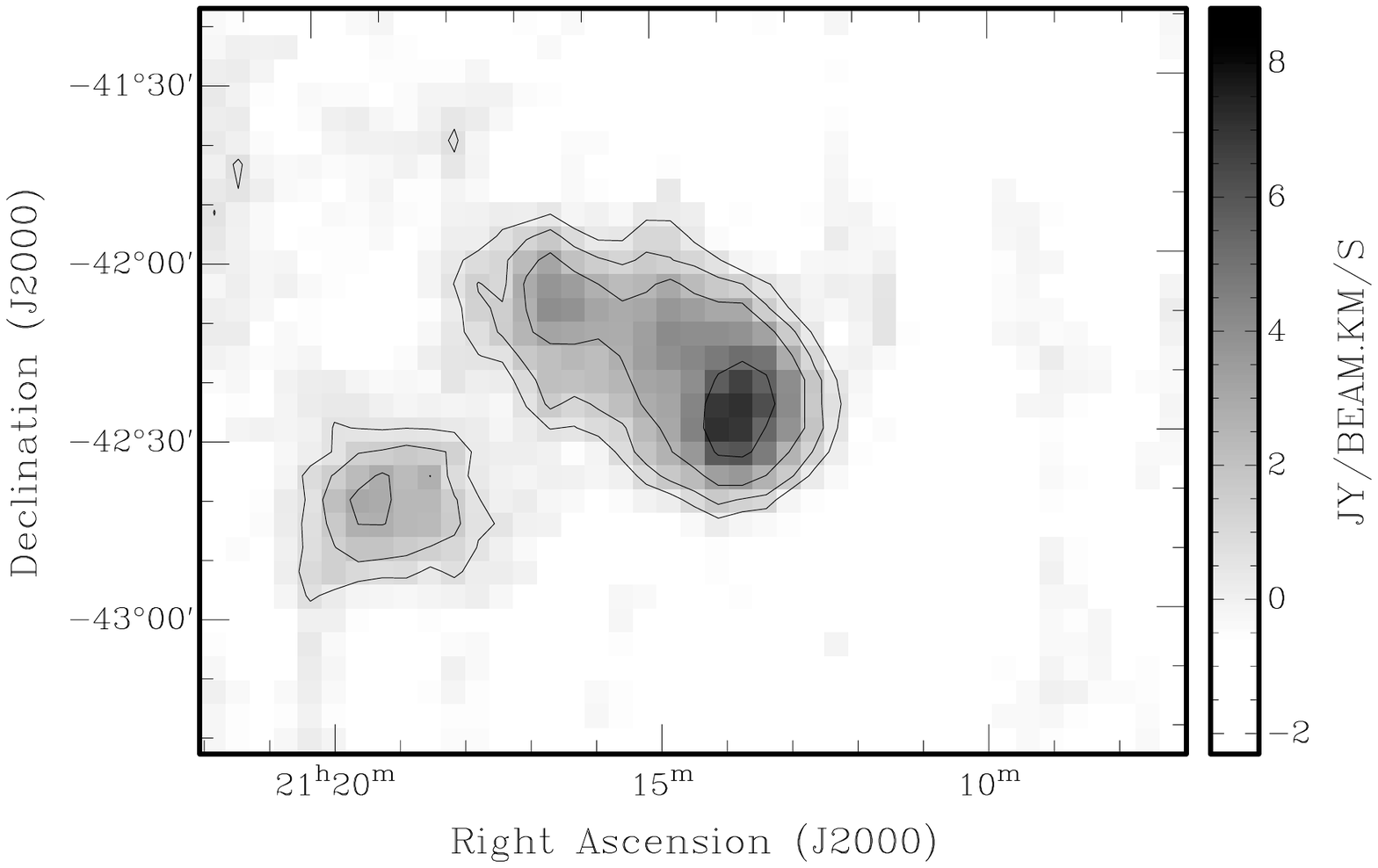}
\includegraphics[height=2.5in]{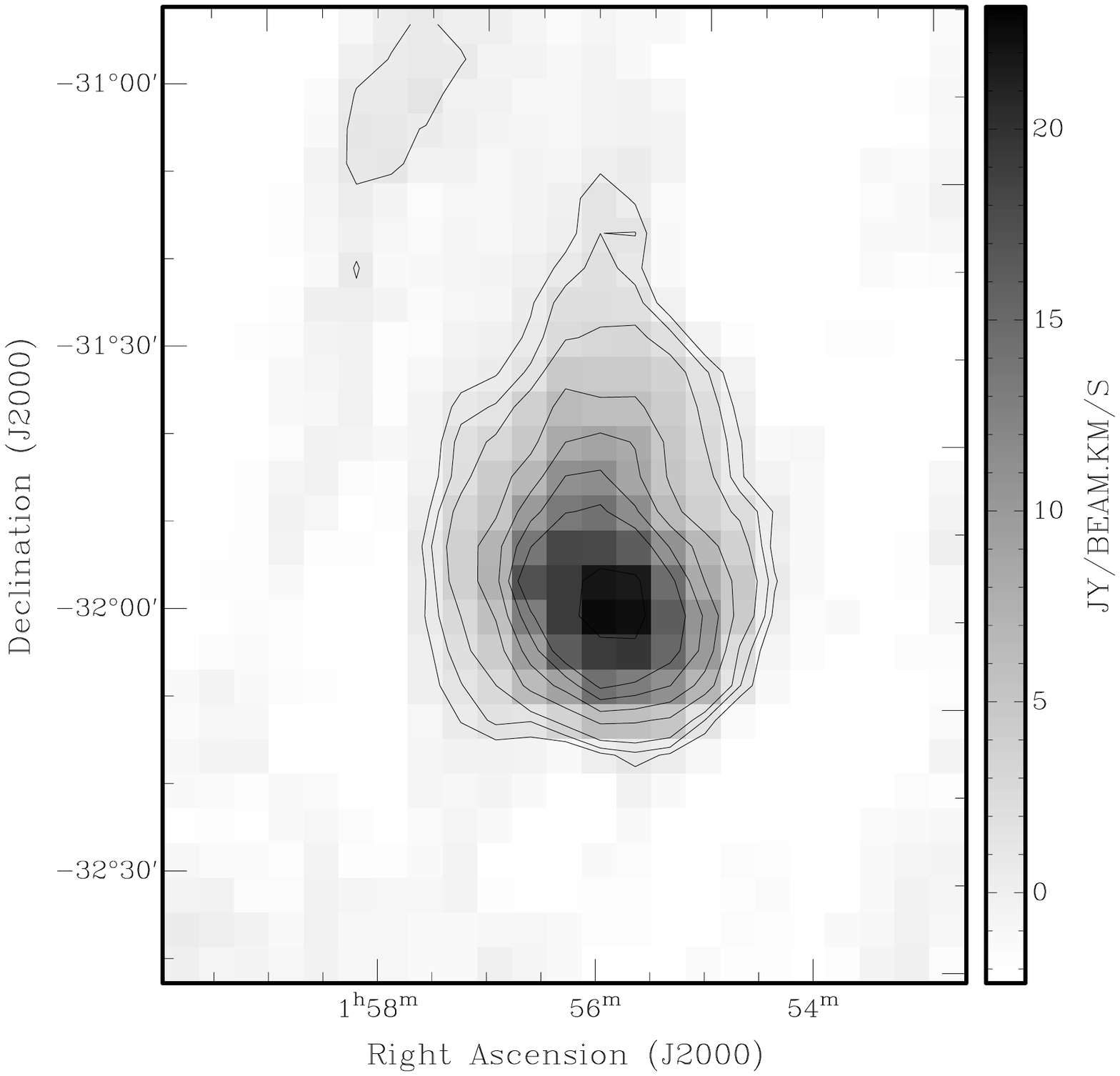}
\includegraphics[height=2.5in]{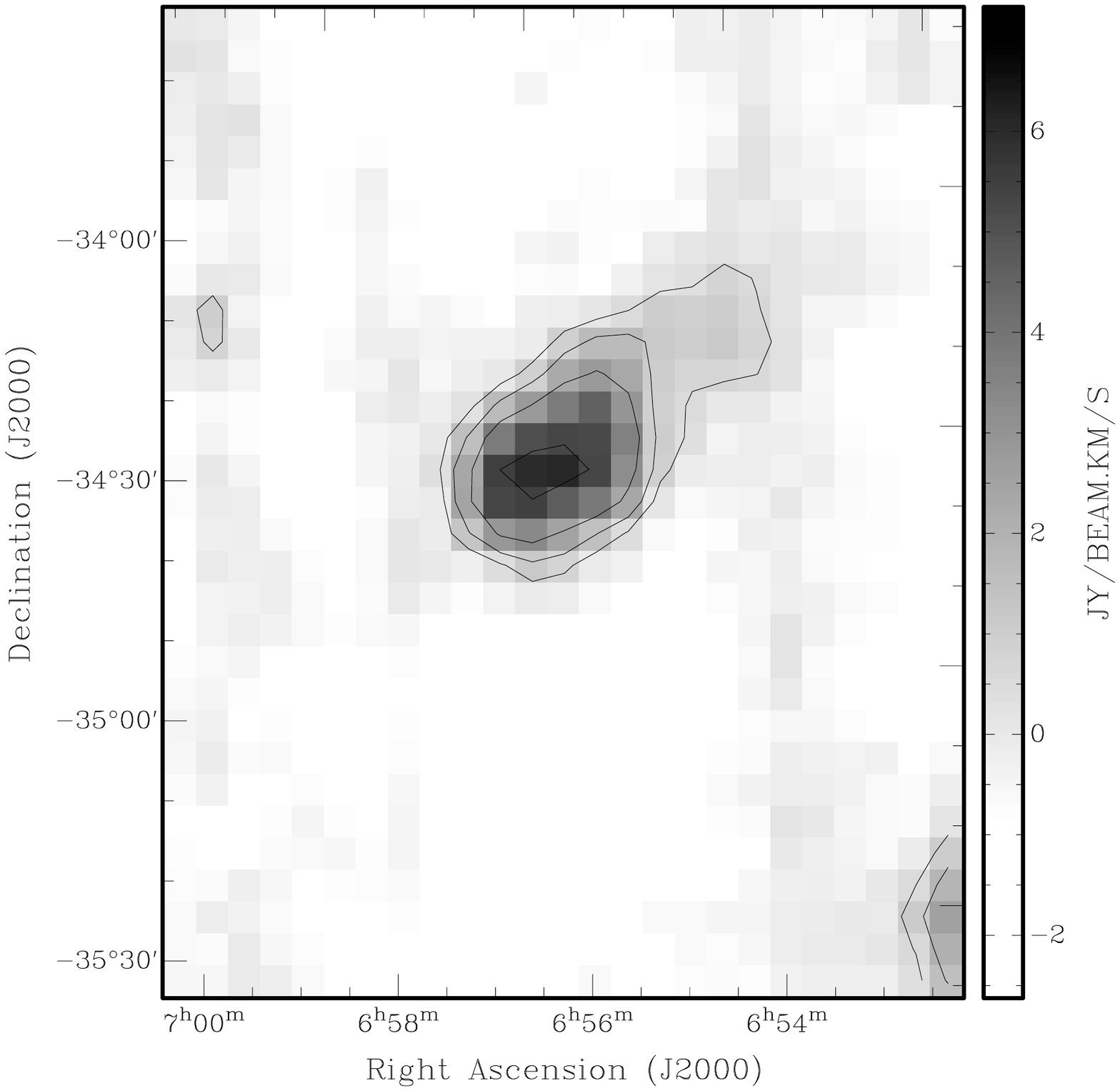}
\vspace{-0.8in}
\includegraphics[height=2.5in]{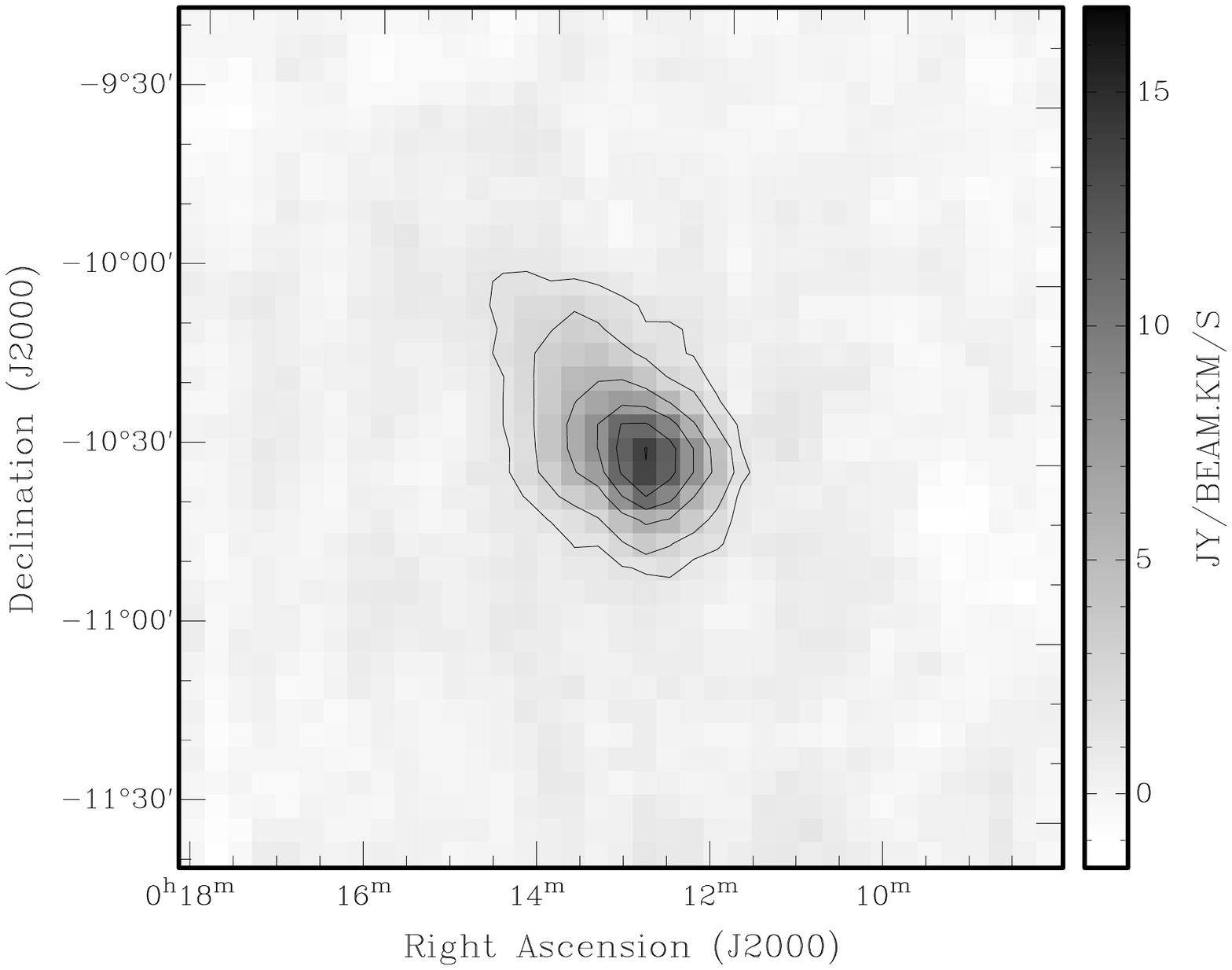}
\includegraphics[height=2.5in]{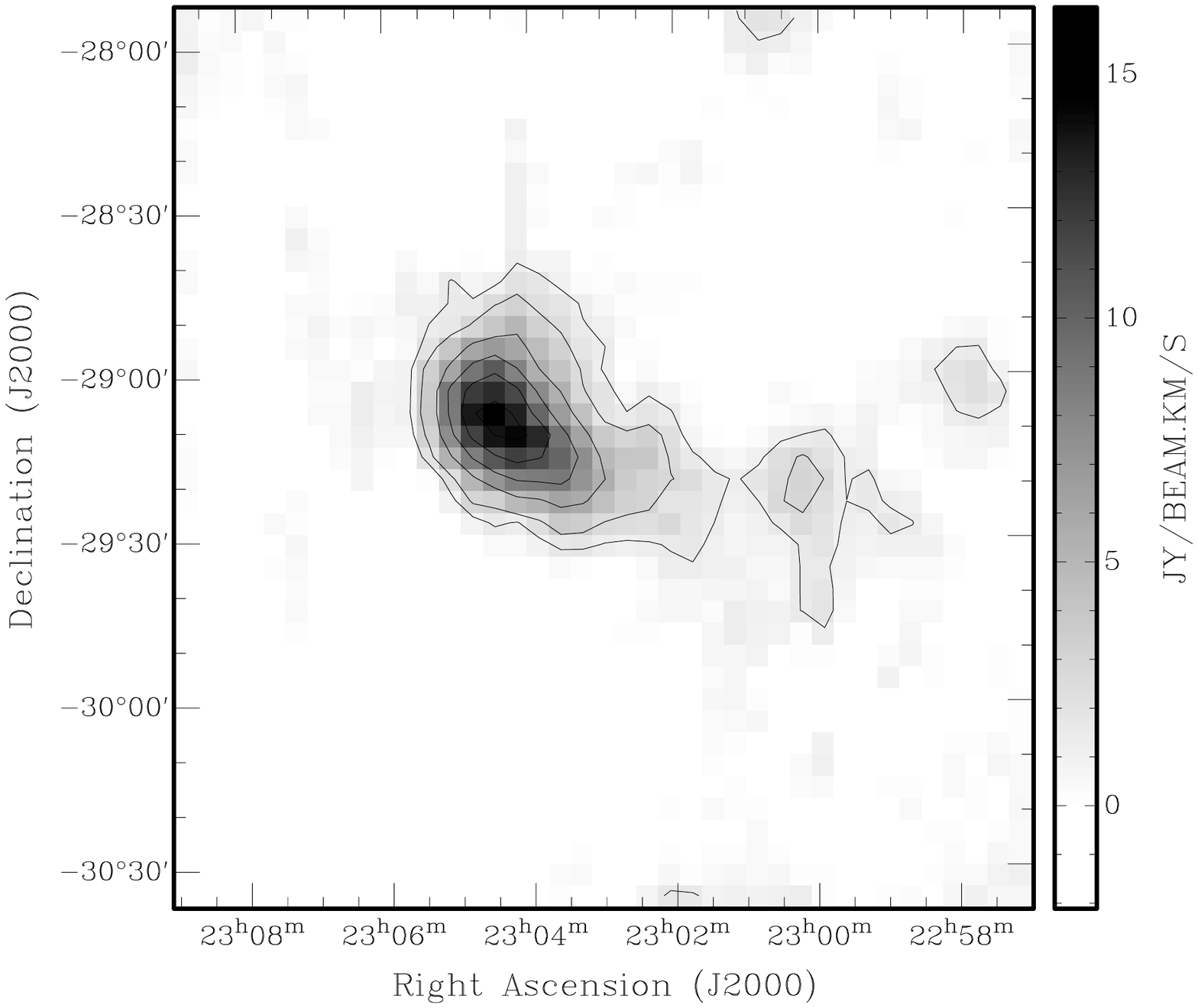}
\includegraphics[height=2.5in]{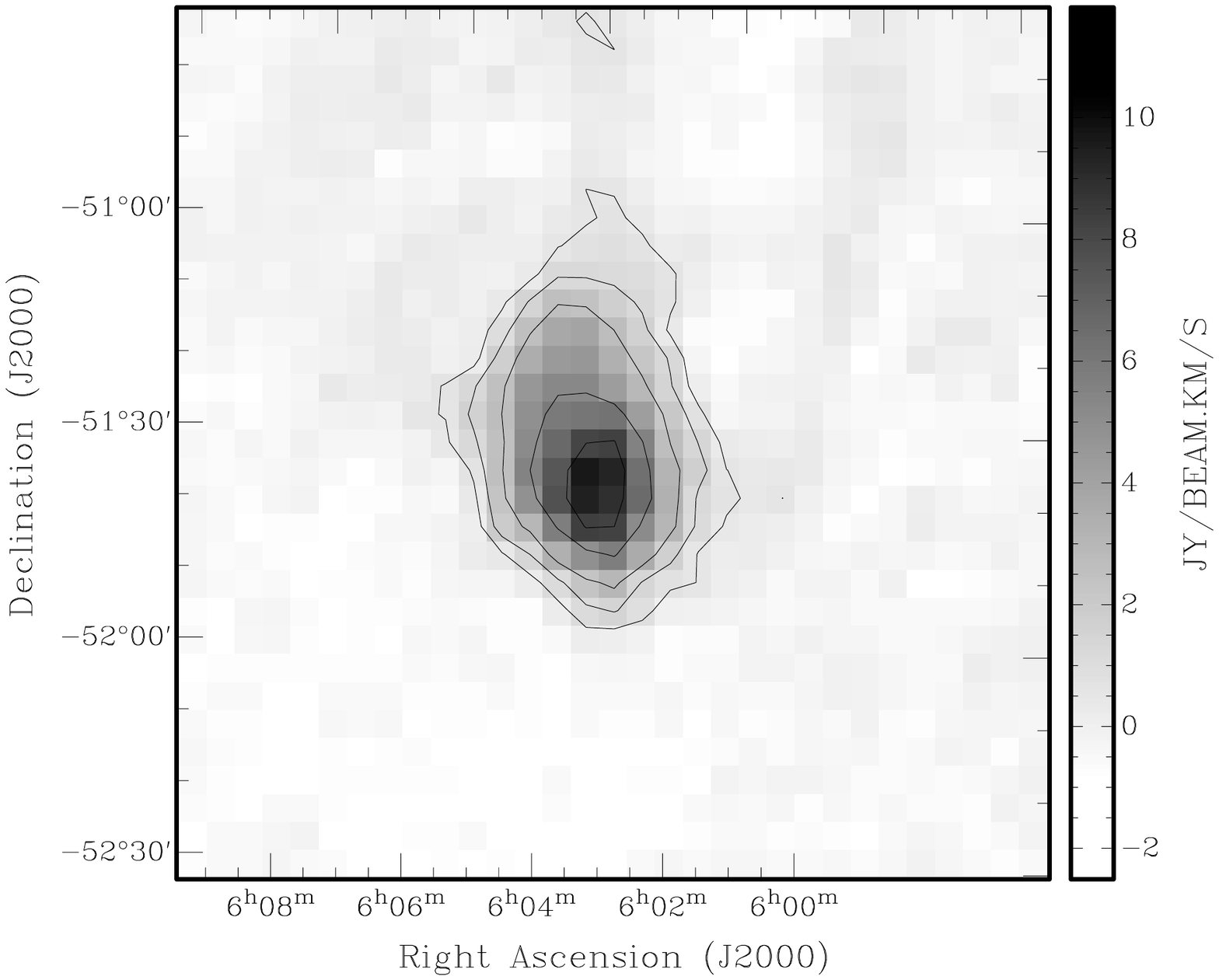}
\end{center}
\caption{Integrated intensity maps of a sample of the 116 head-tail clouds found in the HIPASS data. (P02 cloud \#'s: 1303, 169, 1893, 1986, 700, 759, 501, 210, 933.) The contours are 2 ($5\sigma$), 4, 8, 12, 16, 20, 30 and 40 $\times 10^{18}$ cm$^{-2}$.  See the electronic version for images of all of the head-tail clouds. \label{headtail}}
\end{figure}

\begin{figure}[h!]
\begin{center}
\vspace{-0.4in}
\includegraphics[height=2.5in]{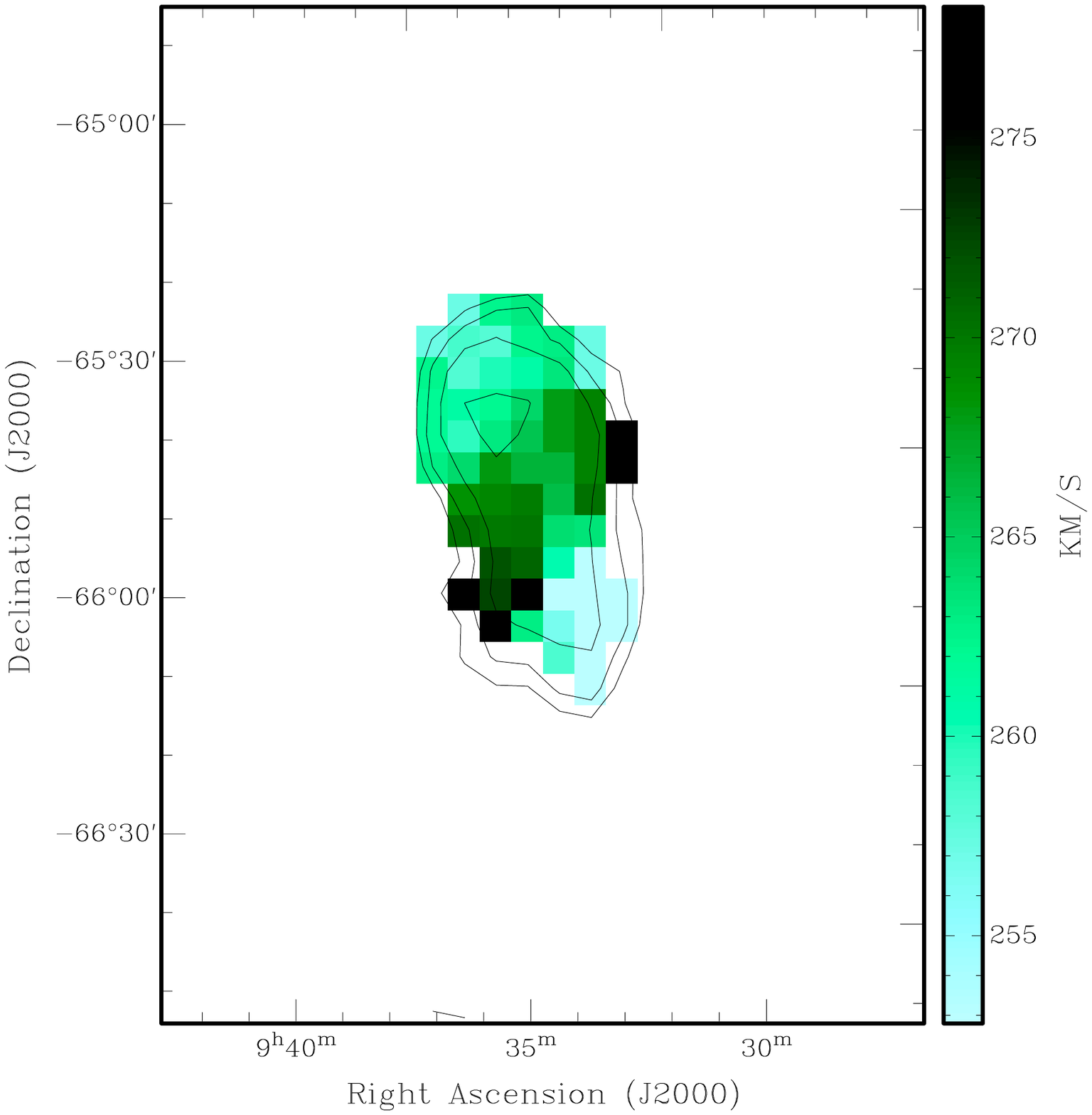}
\vspace{-0.8in}
\includegraphics[height=2.5in]{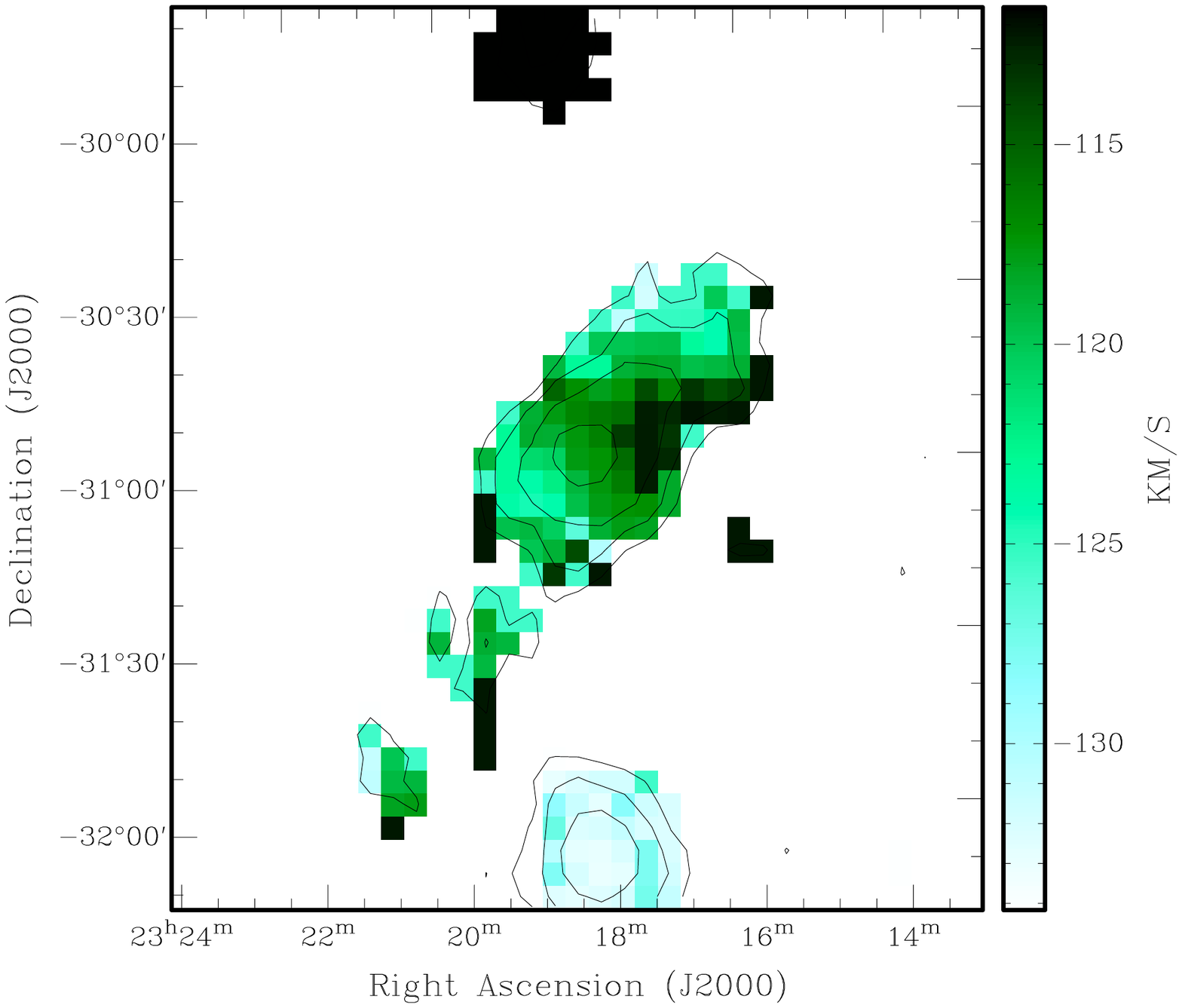}
\includegraphics[height=2.5in]{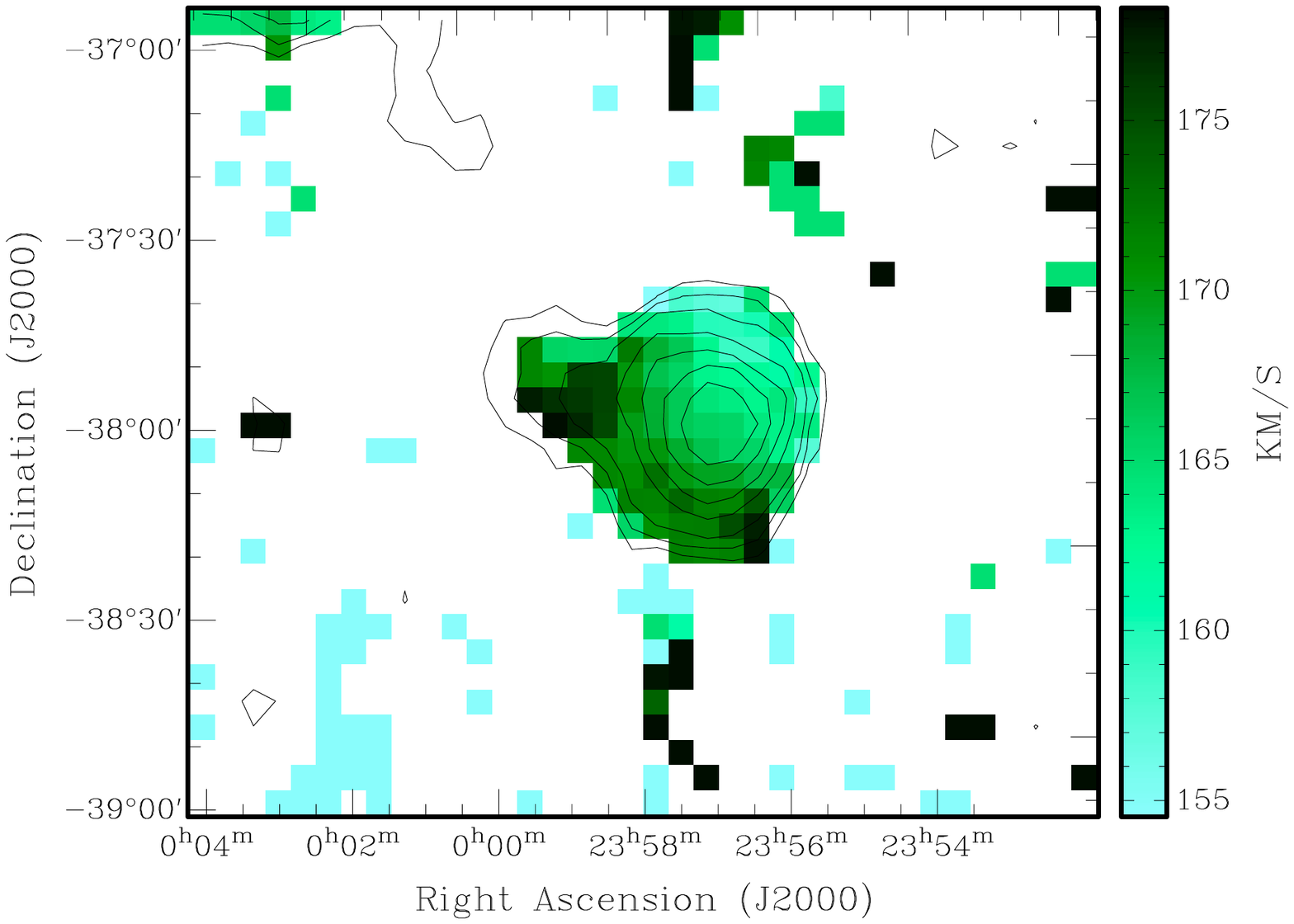}
\vspace{-1in}
\includegraphics[height=2.5in]{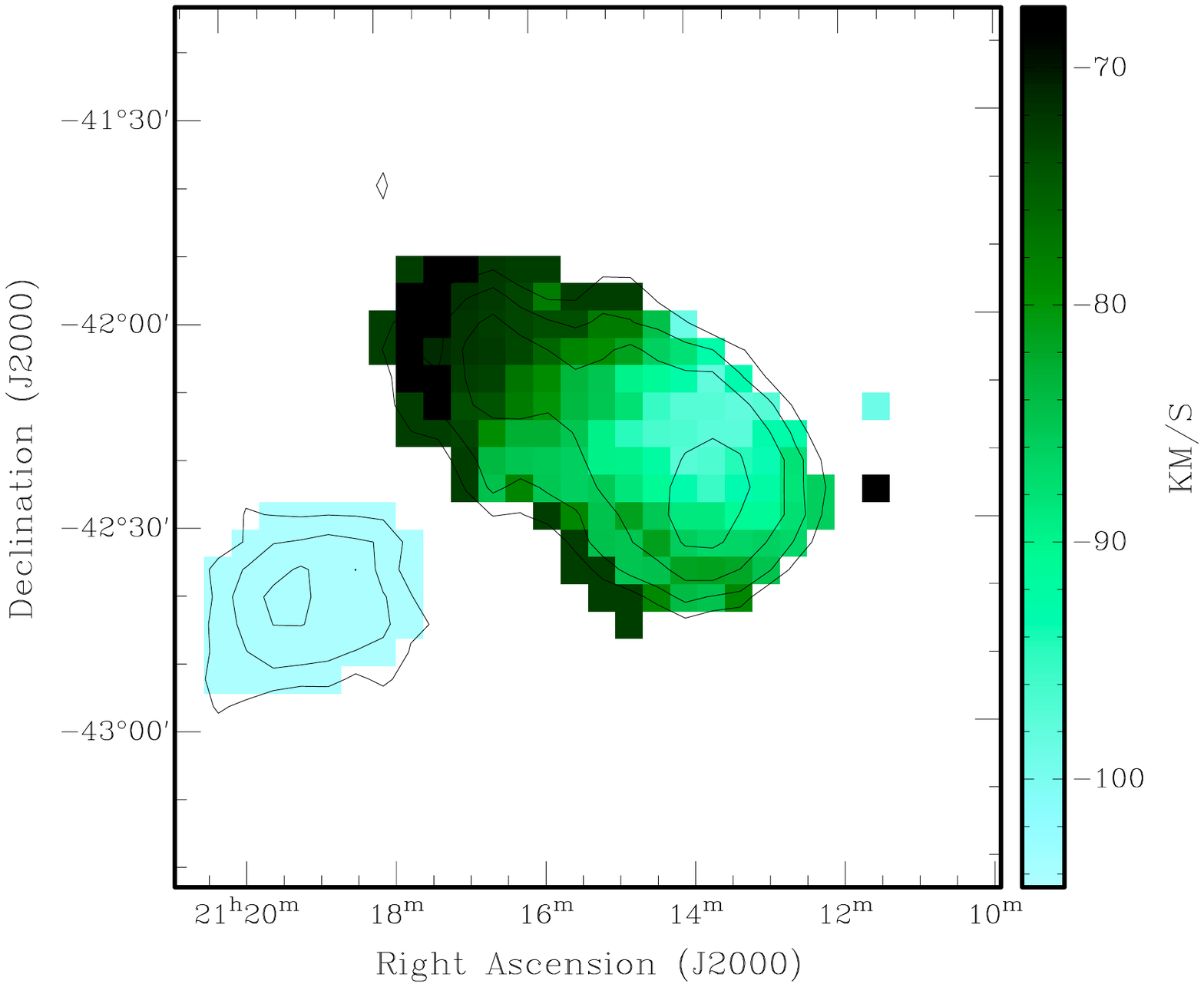}
\includegraphics[height=2.5in]{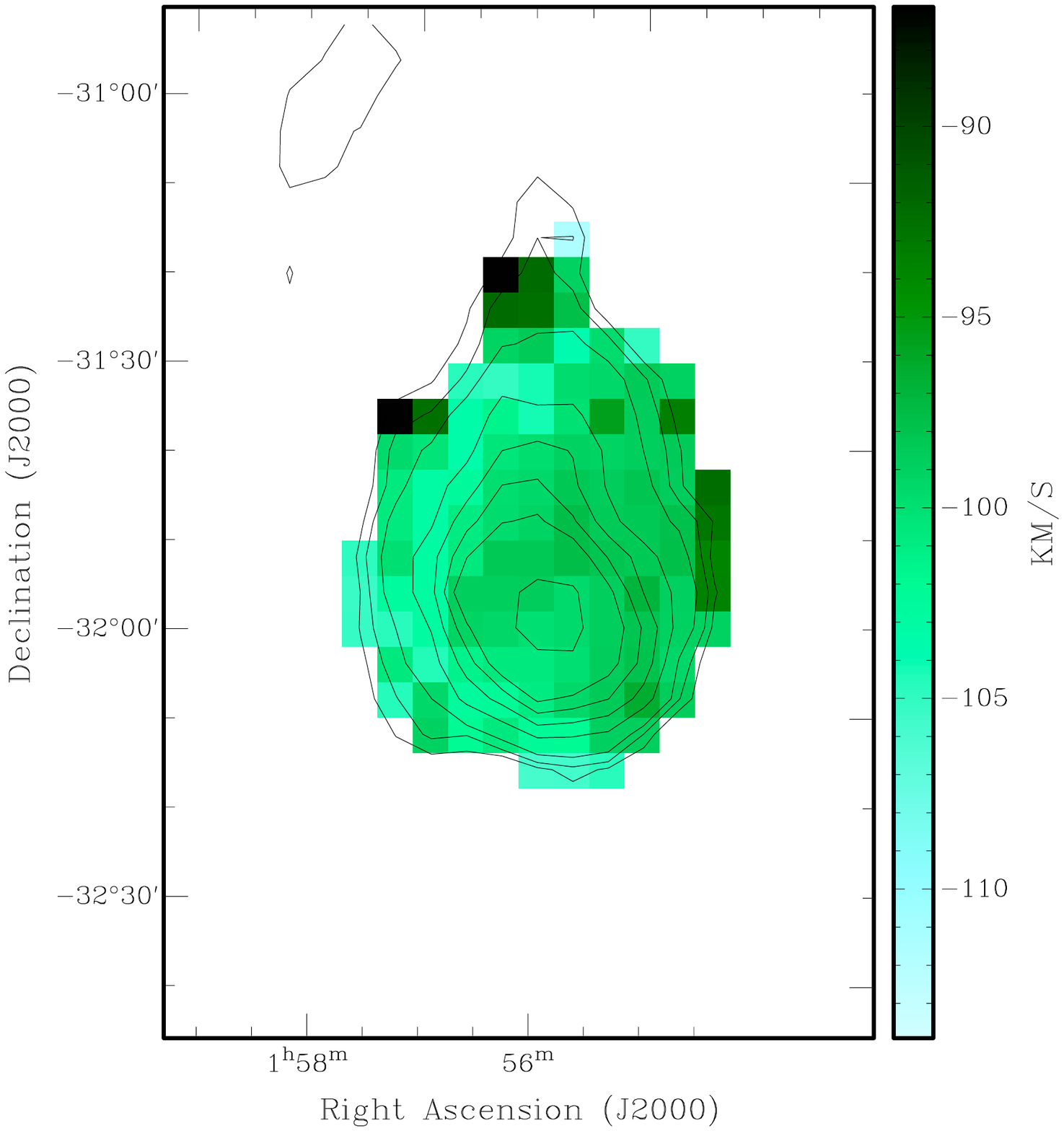}
\includegraphics[height=2.5in]{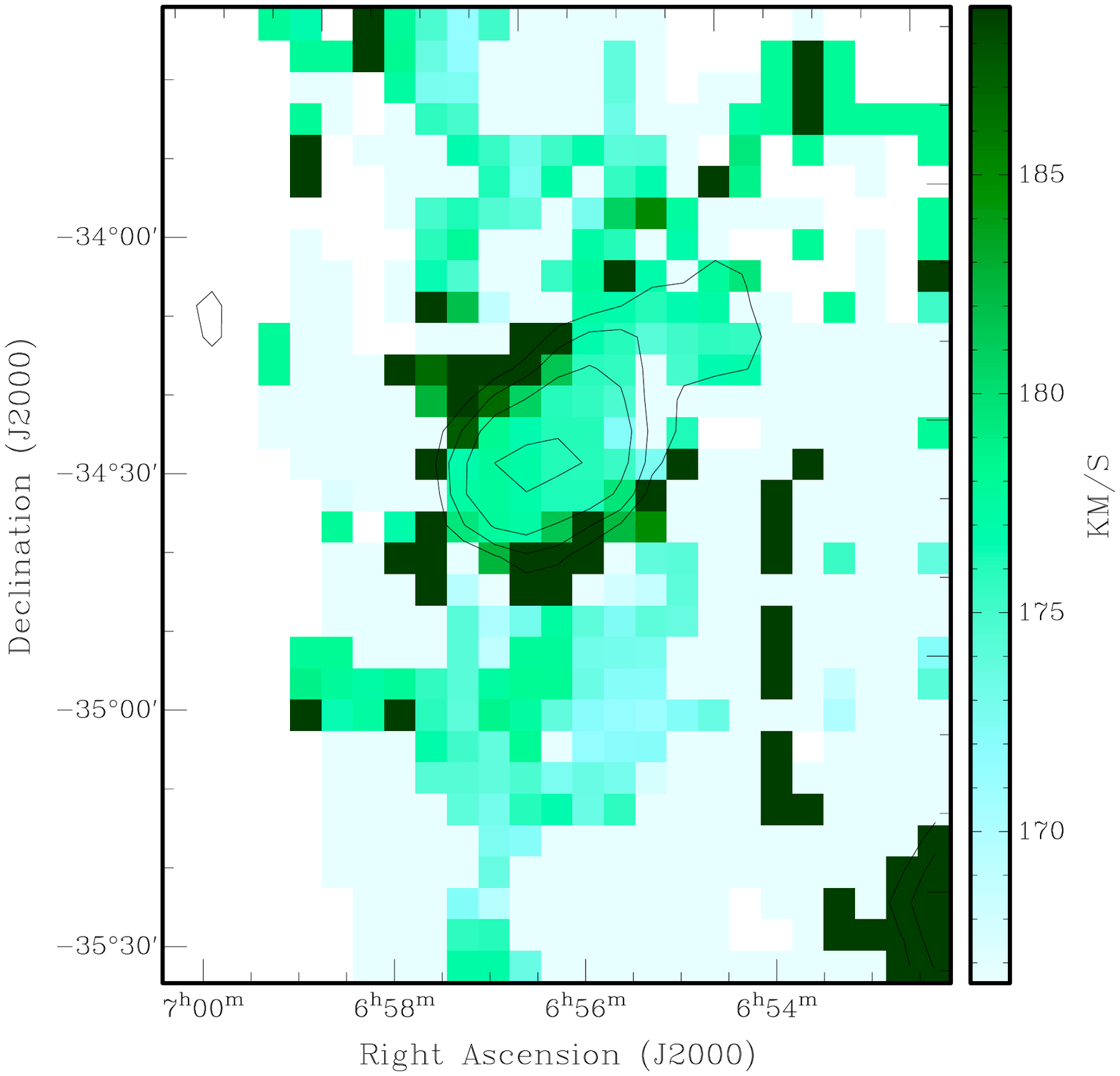}
\vspace{-0.7in}
\includegraphics[height=2.5in]{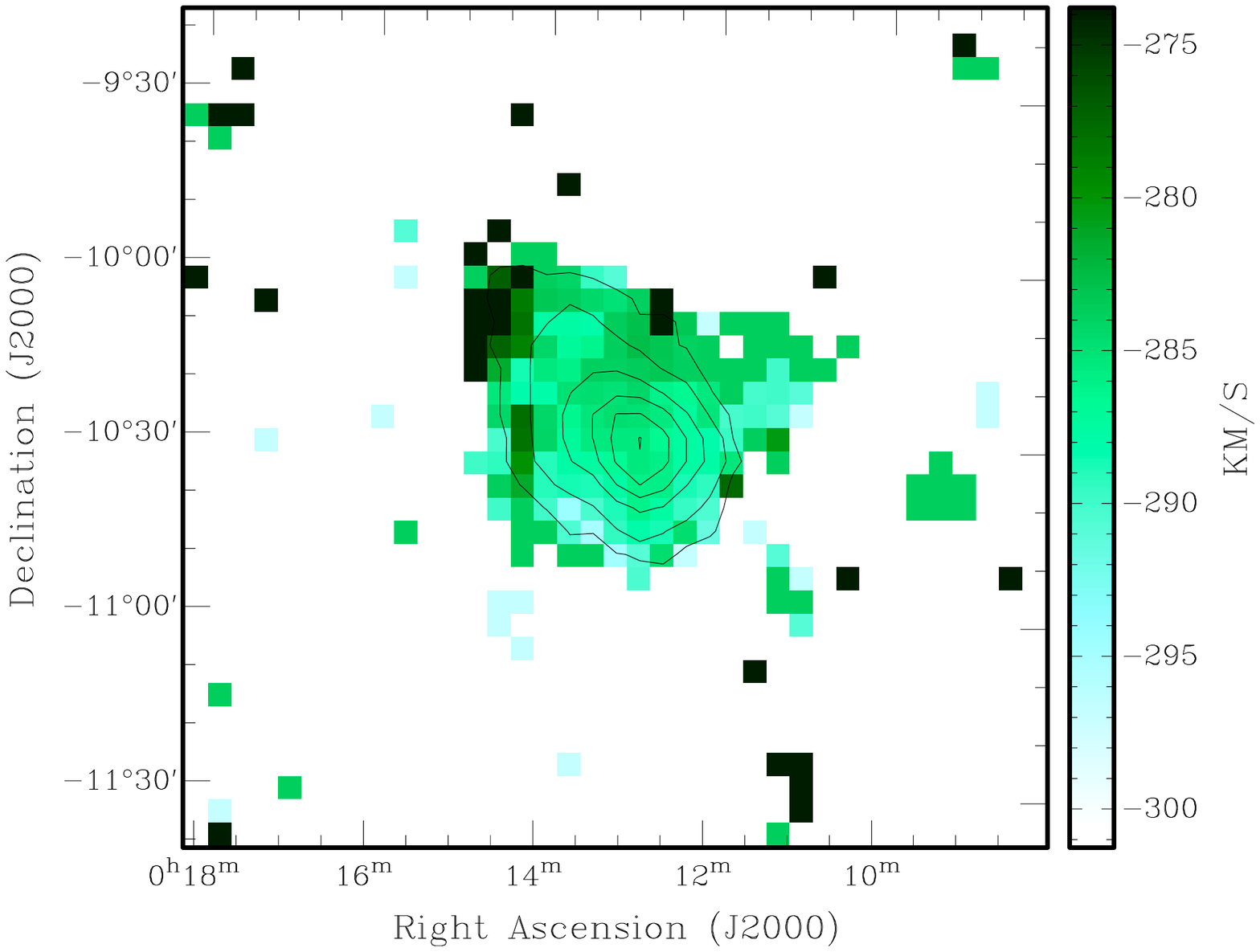}
\includegraphics[height=2.5in]{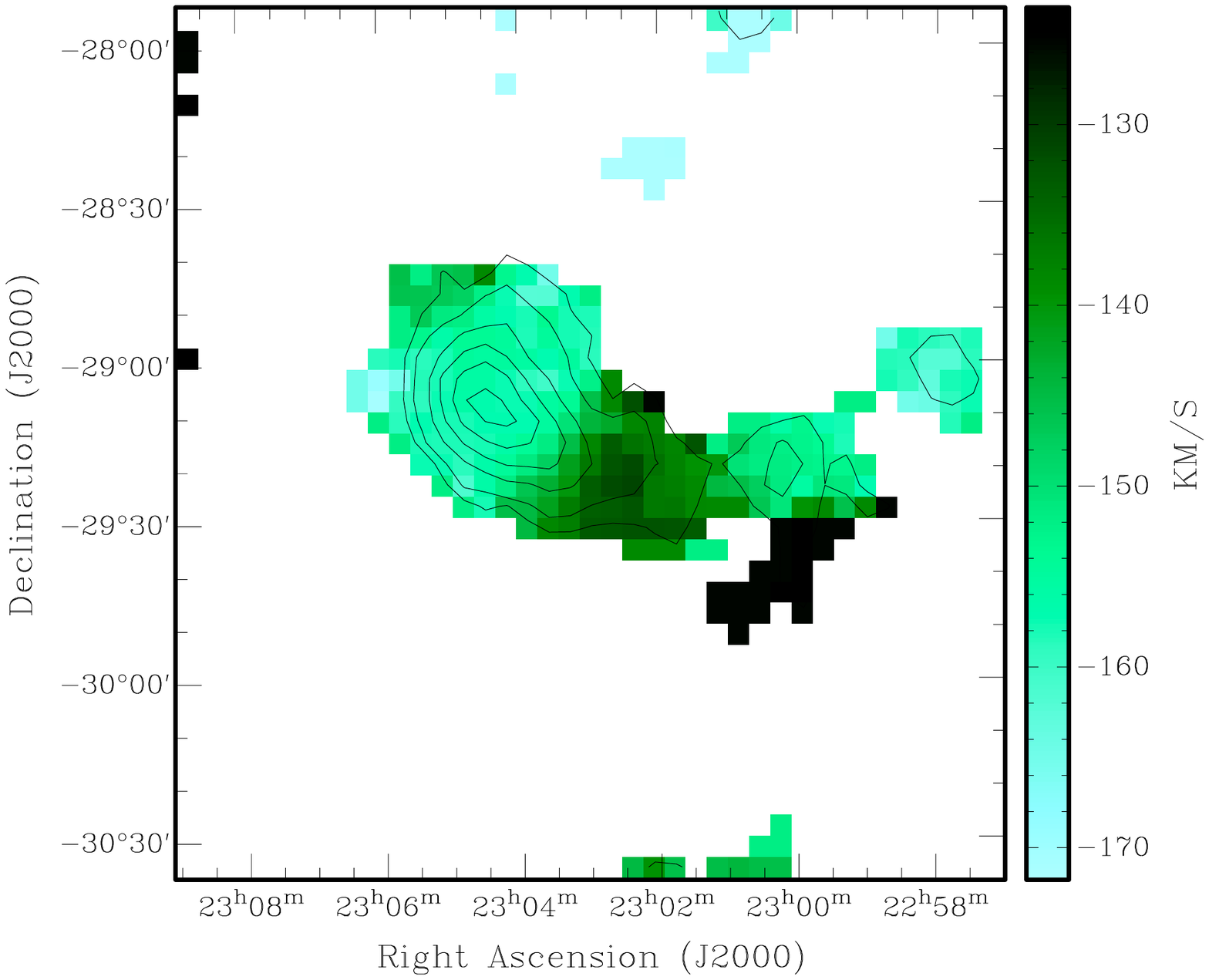}
\includegraphics[height=2.5in]{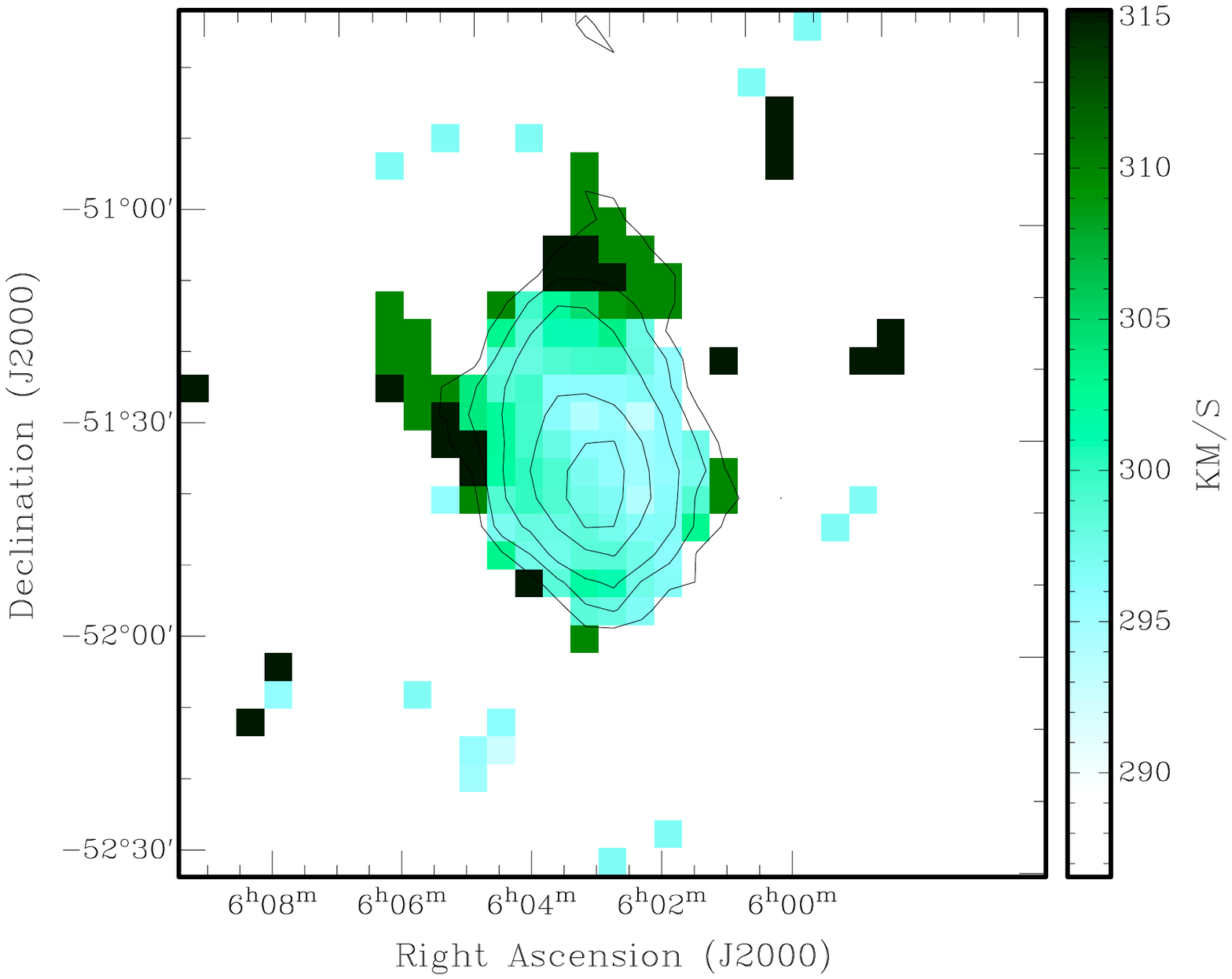}
\end{center}
\caption{The velocity information (LSR) for the same clouds shown in Fig.~\ref{headtail}, with the same contours and
the velocity range shown on the right of each cloud. As noted in the text, the velocity maps were
not used as a strong constraint when defining a head-tail cloud.}
 \label{headtailvel}
\end{figure}

The HIPASS HVCs were cataloged using a friend-of-friend
algorithm described in \cite{deheij02} and P02.
The HVCs were classified into several categories depending on their size
and isolation.   CHVCs, or compact HVCs, are those clouds that 
are less than 2\deg\ in diameter and are isolated from any detectable extended
emission.
:HVCs are clouds that also fit the size criteria of $<2$\deg, but could not be
unambiguously defined to meet the isolation criteria.  The remainder of the large and extended clouds in
the HIPASS dataset are referred to simply as HVCs unless a cloud shows a connection to gas at $|$V$_{\rm dev}| < 60$ \kms~ and then
it is put into a separate catalog and called an XHVC.   Given the HVCs (and XHVCs) in this catalog have large, extended structure that is
generally more complex, we focused on examining the 179 CHVCs and 159 :HVCs for head-tail structure.   
Head-tail clouds were identified by examining the integrated intensity and velocity maps of the CHVCs and :HVCs 
for signatures of a column density and velocity gradient from head to tail.  Given the superior spatial resolution, 
but poor velocity resolution of our data compared to other surveys \citep{bruens00,kalberla05}, the primary criteria for defining a head-tail cloud was the presence of a relatively compact head, trailed by a lower column density, usually narrower tail.  The head also has compressed contours at the opposite end of the tail in the majority of the cases.  
42\% of the CHVCs and 27\% of the :HVCs were defined as head-tail clouds, for a total of 116 head-tail clouds.  The smaller percentage of :HVCs may be partially due to those clouds not being as cleanly defined as distinct clouds from which head-tail structure can be easily identified.
A sample of these clouds is shown in Figures~\ref{headtail} and \ref{headtailvel}.  The integrated intensity maps of the entire sample of head-tail clouds are presented in the electronic version of the journal.

\section{Results}

The head-tail clouds are defined as clouds that have a relatively compact head, trailed 
by a lower column density tail (see Figure~\ref{headtail} for examples).  Many of the clouds show signatures of head compression at the end opposite the tail, as evident from closer contours on this side.  The length and structure of the tail varies substantially, most likely representing a combination of the evolutionary state of the cloud, the viewing angle, and the original structure of the cloud (see \S4). 
The tail generally shows an overall decrease in column density and width
as it gets further from the head of the cloud.  A kink in the tail or an additional clump of gas (of lesser flux
than the main head) is also seen in some clouds.   

\begin{figure}[h!]
\vspace{-0.05in}
\begin{center}
\includegraphics[height=4in, angle=90]{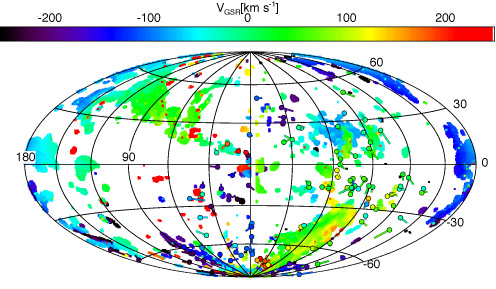}
\end{center}
\vspace{-0.2in}
\caption{The distribution of HIPASS head-tail clouds coded by their GSR velocity on
top of the distribution of HVCs as mapped by the Leiden/Argentine/Bonn (LAB) Survey (Westmeier 2007; Kalberla et al. 2005).  The direction and length of the tail is indicated by the line trailing the point, with the length magnified by a factor of 4. }
 \label{ait}
\end{figure}

\begin{figure}[h!]
\begin{center}
\includegraphics[height=4.5in, angle=-90]{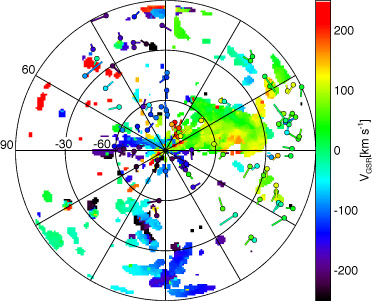}
\end{center}
\vspace{-0.2in}
\caption{The distribution of HIPASS head-tail clouds coded by their GSR velocity on
top of the distribution of HVCs as mapped by the LAB Survey (Westmeier 2007; Kalberla et al. 2005) in polar projection for $b < 0$\deg.  The direction of the tail is indicated by the line trailing the point, with the length magnified by a factor of 4. }
  \label{polar}
\end{figure}

\begin{figure}[ht!]
\vspace{-0.05in}
\begin{center}
\includegraphics[height=3.9in, angle=90]{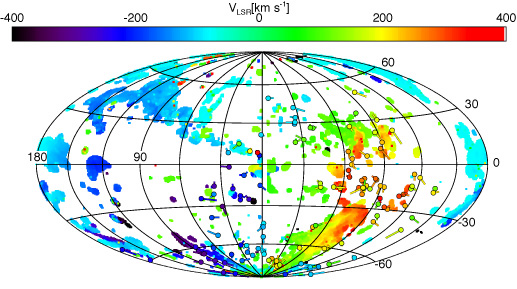}
\end{center}
\vspace{-0.2in}
\caption{The distribution of HIPASS head-tail clouds coded by their LSR velocity on
top of the distribution of HVCs as mapped by the LAB Survey (Westmeier 2007; Kalberla et al. 2005).  The direction and length of the tail is indicated by the line trailing the point, with the length magnified by a factor of 4.  }
 \label{ait2}
\end{figure}

\begin{figure}[ht!]
\begin{center}
\includegraphics[height=4.5in,angle=-90]{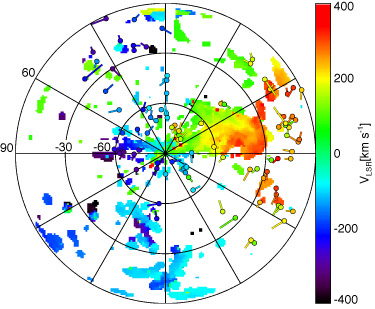}
\end{center}
\vspace{-0.2in}
\caption{The distribution of HIPASS head-tail clouds coded by their LSR velocity on
top of the distribution of HVCs as mapped by the LAB Survey (Westmeier 2007; Kalberla et al. 2005) in polar projection for $b < 0$\deg.  The direction of the tail is indicated by the line trailing the point, with the length magnified by a factor of 4. }
  \label{polar2}
\end{figure}

The velocity structure of the head-tail clouds varies (see Figure~\ref{headtailvel}) and, as noted in \S 2, our ability to analyze this feature of the clouds is limited due the velocity resolution of the HIPASS data and the typical linewidth of HVCs being $< 30$\kms.  
For 44 (37\%) of the head-tail clouds a clear velocity gradient could be defined, with the difference in central velocity between the head and tail $>$ 20\kms.   If the absolute value of the velocity of the tail is less than that of the head, the tail can be defined as lagging the head.  For the 25/44 head-tail clouds with negative LSR velocities, all but one has the tail lagging the head.  This is also the case when these clouds are considered in the GSR frame.   On the other hand, 16 of the 19/44 positive V$_{\rm LSR}$ clouds with a definable gradient show the opposite effect, with the tail at a higher positive velocity than the head.  When the GSR frame is examined, the number is 13 instead of 16.  These positive velocity clouds are largely found in the region leading the Magellanic System (see below and Figures~\ref{ait}-\ref{polar2}).  It is relevant to note that the unmeasured velocity component (tangential to the line of sight) could dominate the motion of the head-tail clouds, as the presence of the spatial head-tail structure immediately indicates they are moving tangential to the line of sight at some level.

 Figures~\ref{ait} - \ref{polar2} show the distribution
of HVCs on the sky and their average velocities using the LAB data (Westmeier 2007\nocite{westmeier07t}) and the positions and velocities of the HIPASS head-tail clouds.  The LAB HVC map was created by subtracting out a three-dimensional model of the Milky Way's disk gas \citep{kalberla07} which differs slightly from how the CHVCs and :HVCs were defined in P02.   This together with the decrease in sensitivity and bigger beam of the LAB data results in the majority of the HIPASS head-tail clouds not being on top of an LAB cloud in this image (the latter being the larger effect).  
By the original definition of a CHVC or :HVC in P02 the head-tail clouds are relatively isolated from extended emission or larger complexes; however, as shown in Figures~\ref{ait}-\ref{polar2} and noted in P02, they are generally grouped and/or found in the vicinity of larger complexes of clouds\nocite{kalberla05}.

\begin{figure}
\vspace{-0.3in}
\includegraphics[height=3.5in, angle=0]{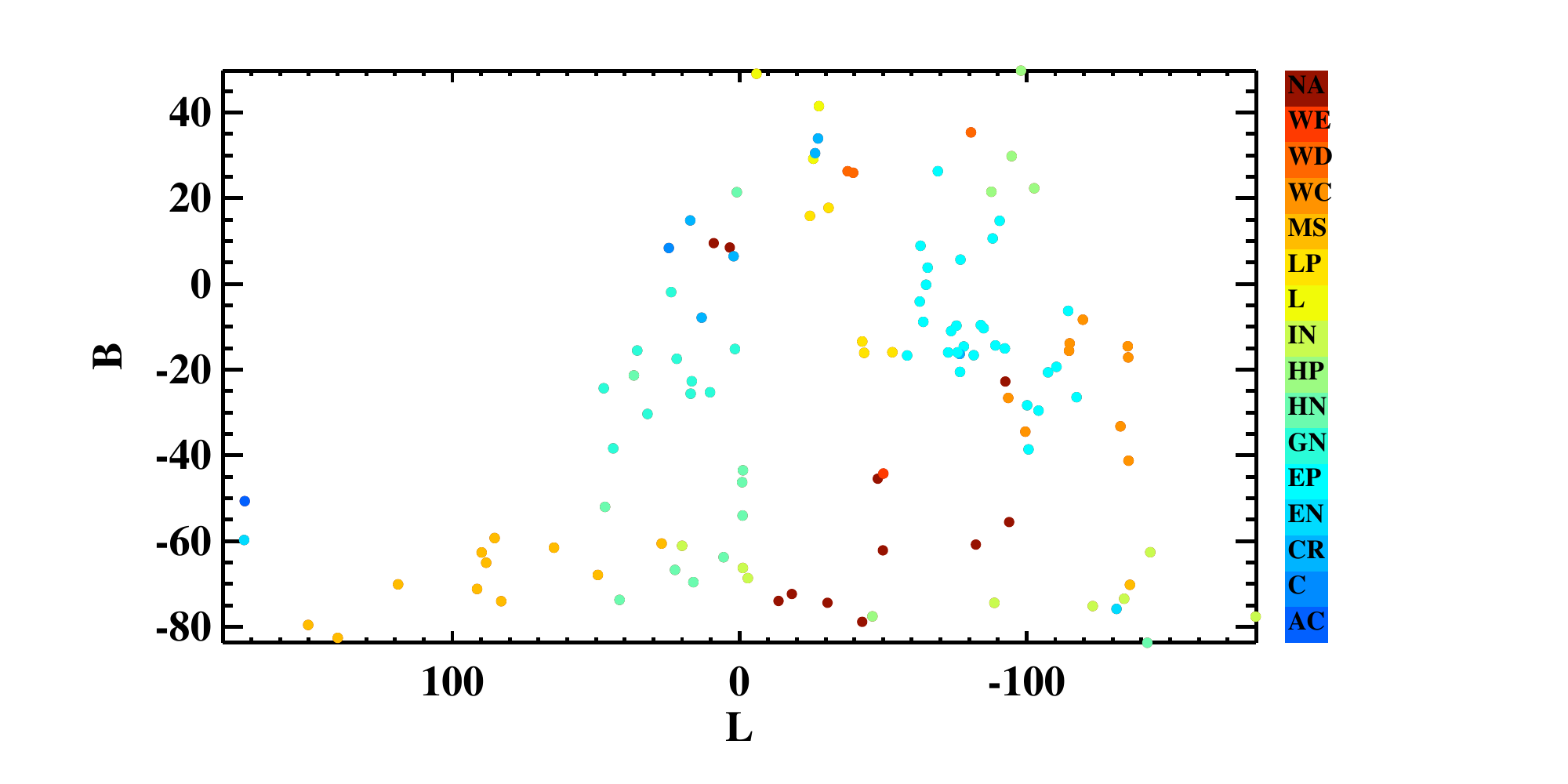}
\vspace{-0.3in}
\caption{The distribution of HIPASS head-tail clouds in Galactic coordinates with the color representing the name of the HVC complex that can be associated with the head-tail cloud (see \S3).    \label{complb}}
\end{figure}

\begin{figure}
\vspace{-0.3in}
\includegraphics[height=3.5in, angle=0]{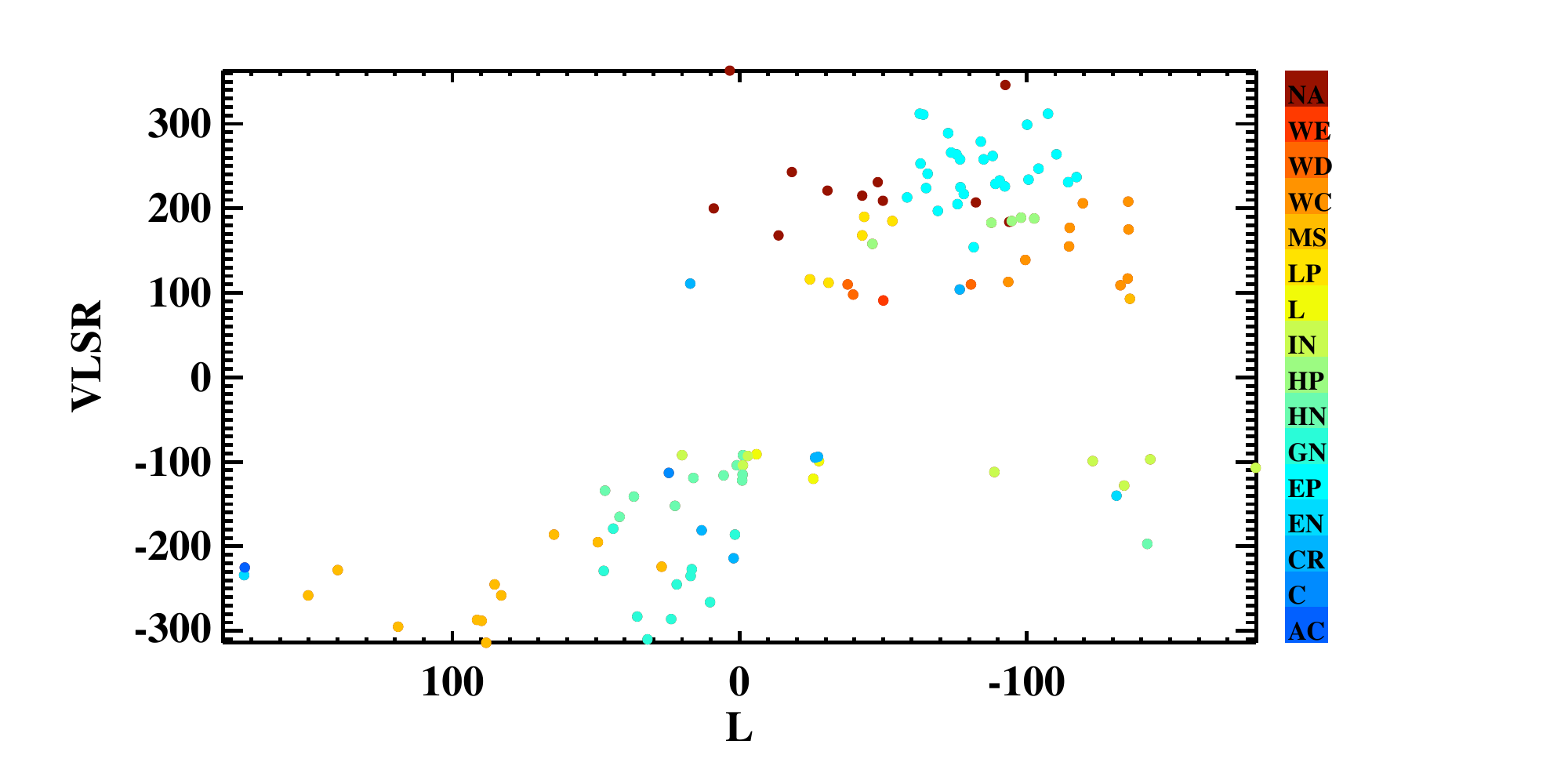}
\vspace{-0.3in}
\caption{The distribution of HIPASS head-tail clouds in Galactic longitude vs. V$_{\rm LSR}$ with the color representing the HVC complex that can be associated with the head-tail cloud (see \S3). \label{complv}}
\end{figure}

To further investigate the association of the head-tail clouds with large HVC complexes we used the criteria developed by \cite{peek08} to group simulated halo clouds in a similar fashion to how the HVCs have been associated into complexes by \cite{wakker91} (hereafter WvW91).  The criteria is 
\begin{equation}\label{assoc}
D = \sqrt{\Theta^2 + (0.5~\degr/{\rm km~s}^{-1})^2\left(\delta v\right)^2},
\end{equation}
\noindent where $\Theta$ is the angular distance in degrees between a head-tail cloud and a HVC complex cloud in WvW91, and $\delta v$ is the corresponding difference in velocity.  The data used to make the WvW91 catalog are of similar resolution and sensitivity to the data shown in Figures~\ref{ait} - \ref{polar2}.   Head-tail clouds with $D < 25$\deg~ were called associated with the corresponding HVC complex, as this value was found optimal for the clustering criteria in \cite{peek08}.   
The results of the association of head-tail clouds with HVC complexes are shown in Figures~\ref{complb} and \ref{complv}, with the head-tail clouds shaded by the associated WvW91 HVC complex name.  We found only 11 of the 116  head-tail clouds were not associated (labeled NA) with a HVC complex.  If the Magellanic Clouds and Bridge were included in the HVC catalog this number would decrease further, as the majority of the NA clouds are found in this region\footnote{One particularly notable exception is the NA cloud at $l, b, V_{\rm LSR} =$ 3.4\deg, 8.5\deg, 365 \kms~that may be a galaxy not obvious in the optical due to its low Galactic latitude.}.    The complexes labeled EP and HP can be associated with the Leading Arm of the Magellanic System and there are 33 head-tail clouds associated with these complexes.  The majority of the 9 clouds associated with WC also represent leading debris from the Magellanic System.  All of these clouds are visible in Figures~\ref{ait} and \ref{ait2} at $\sim l=210 - 330$\deg, $b> -30$\deg~and V$_{\rm LSR} > 150$\kms~and are described in the context of the Magellanic System in \cite{bruens05}, \cite{putman98} and \cite{bekhti06}.   There are 12 clouds associated with the Magellanic Stream (labeled MS).  The other cloud complexes named are not known to have a distinct origin, although many have distance constraints (see \S4.1).
 
\begin{figure}
\begin{center}
\vspace{-0.01in}
\includegraphics[height=2.3in, angle=0]{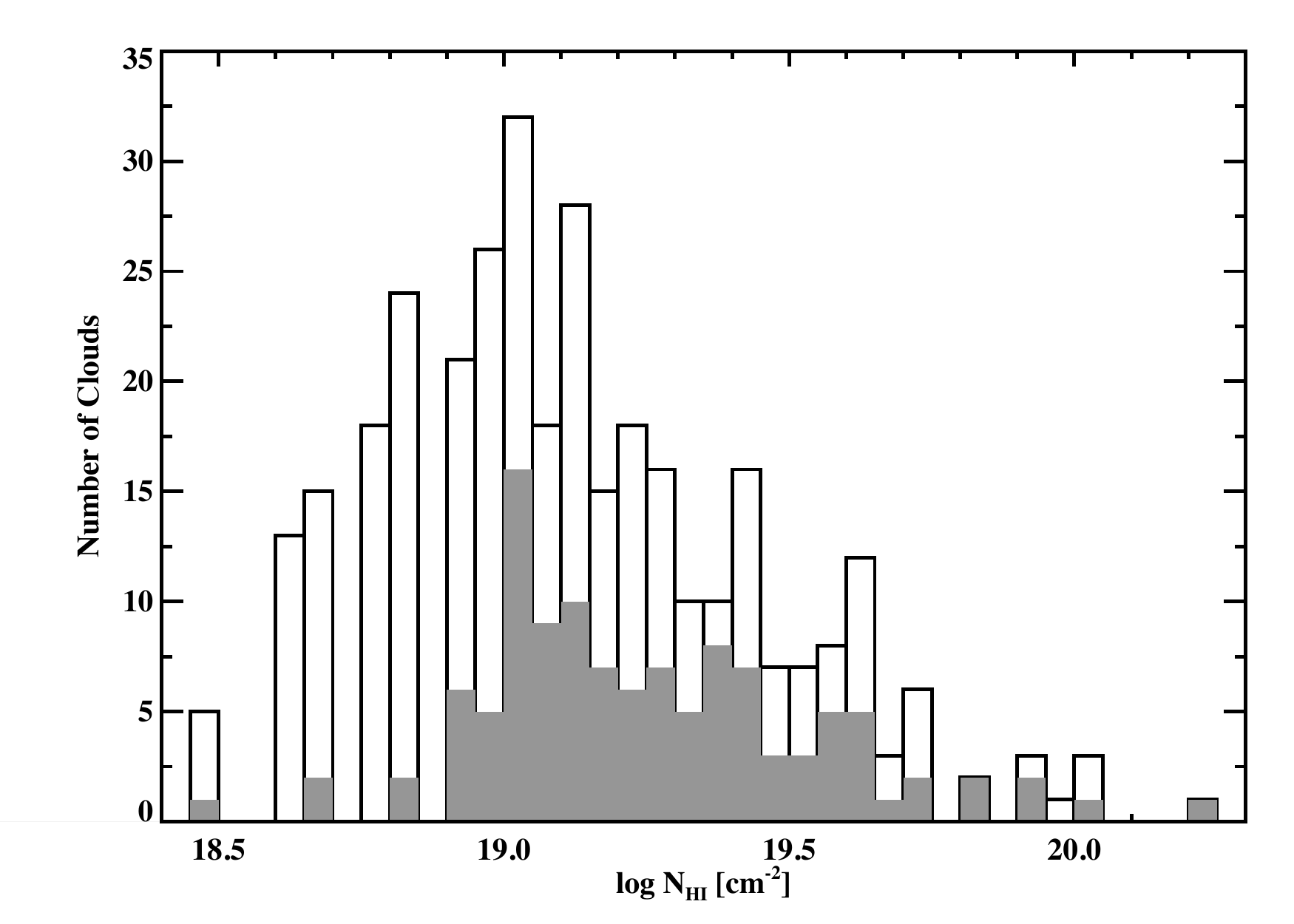}
\includegraphics[height=2.3in, angle=0]{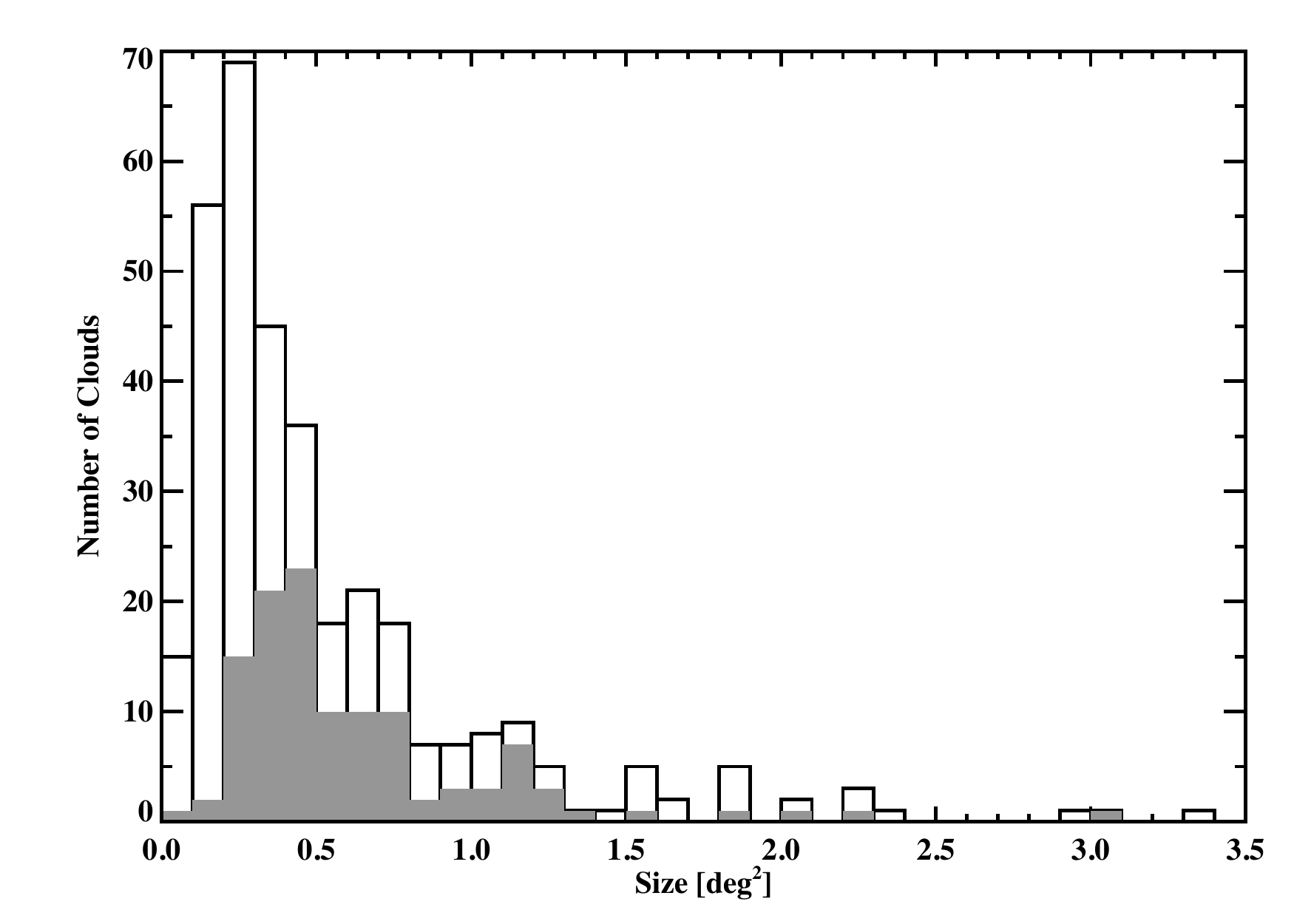}
\hspace{-0.30in}
\vspace{-0.30in}
\includegraphics[height=2.3in, angle=0]{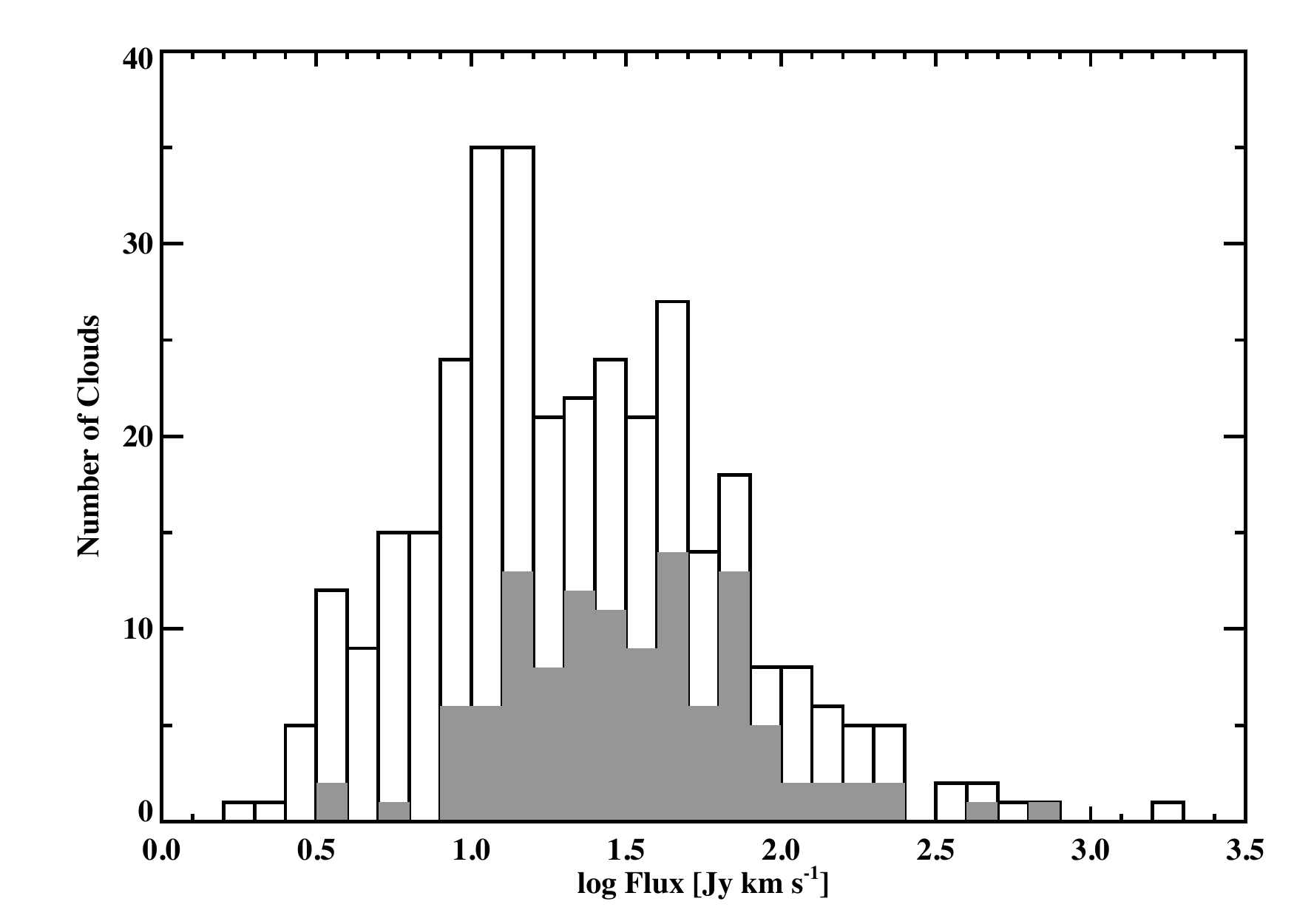}
\includegraphics[height=2.3in, angle=0]{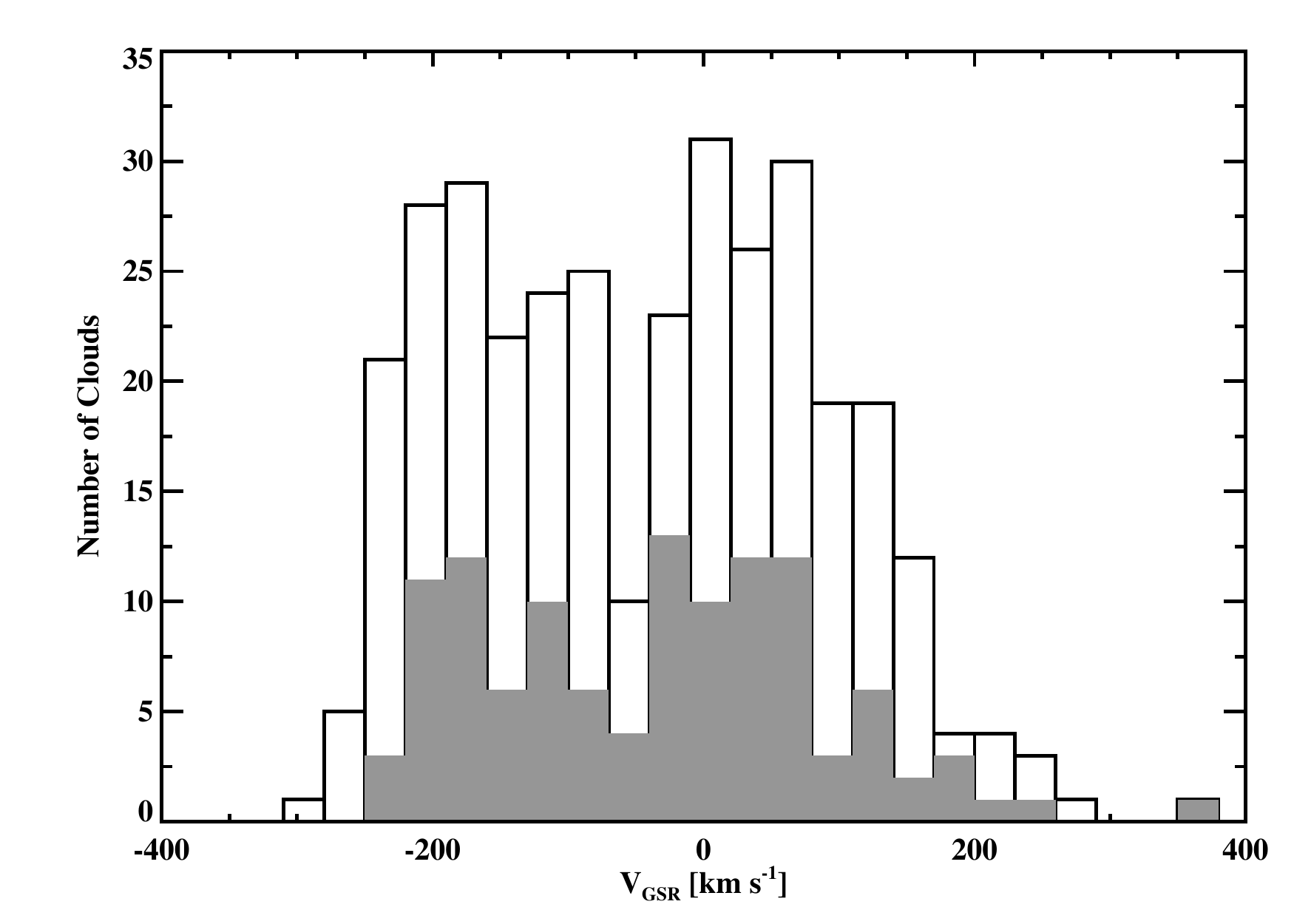}
\end{center}
\caption{The distributions of peak column densities, sizes, total fluxes, and velocities (GSR) for all of the CHVC and :HVC clouds (open histogram), and just the head-tail clouds (shaded).}  \label{props}

\end{figure}

The distributions of peak column densities, sizes, total fluxes, and velocities in the Galactic Standard of Rest frame (V$_{\rm GSR}$) for the head-tail clouds are shown in Figure~\ref{props} and compared to the properties of the HIPASS CHVCs and :HVCs that did not exhibit head-tail structure.    The only distinct difference between the head-tail clouds and the non-head-tail clouds is that the head-tail clouds are shifted towards the higher ends in the distributions of peak column density, total flux, and size.  This may be expected as clouds with lower peak column densities will have an even lower column density tail that may be beyond our sensitivity and small clouds may have a tail blended with the head due to our resolution limits.
The head-tail clouds have typical peak column densities of 
N$_{HI} = 10^{18.9-19.7}$ cm$^{-2}$, with the median value at $10^{19.2}$ cm$^{-2}$ and a similar mean value.
The total fluxes of the head-tail clouds are typically between $10-100$ Jy \kms, with the median value at 30 Jy~\kms.  At 10 kpc, 30 Jy~\kms~ is equivalent to 730 \Msun, and at 60 kpc it is equivalent to 26,000 \Msun (M$_{\rm HI} \propto$ D$^2$).   The P02 catalog of HVCs was assessed to be complete at these flux values.
The size for each cloud is measured in terms of the number of pixels covered by each cloud times the area of a 4\arcmin $\times$ 4\arcmin~pixel and shows how all but one of the clouds has a size $<2$ deg$^{2}$, with the majority $<1$ deg$^{2}$.   
The peaks in these distributions may continue to evolve with higher sensitivity and resolution surveys (e.g., see \cite{hsu11}).  
The GSR velocities of the head-tail clouds are largely $|$V$_{\rm GSR}| < 250$ \kms~and there is no significant difference in the velocity range of the clouds with and
without head-tail structure.
The mean velocity of the HT clouds is V$_{\rm GSR}\sim -37$ \kms~ and the median is $-25$ \kms.  This is suggestive of net infall, but the ability to strongly interpret these values is limited due to only being able to measure the line of sight velocity.  These values are again similar to that of the entire population of HIPASS CHVCs and :HVCs (P02).

We also investigated the ratio of mass in the head verses tail of the cloud.  The flux in the head was determined by fitting an elliptical gaussian to the head of the cloud and inspecting the fit by eye to ensure the fit encompassed as little of the tail as possible.  Since the head and tail are directly connected and likely to be at approximately the same distance, the flux ratio can be discussed directly in terms of mass.
The head-tail clouds almost always contain more mass in their heads than tails, with the most common values of M$_{\rm head}$/M$_{\rm tail}$ between 1 - 2.5.  
The head-tail clouds therefore typically have 50-70\% of their mass in their head, although this has not been corrected for any inclination effects.  Simulations indicate there is not expected to be a large amount of mass in diffuse tail material not detectable by HIPASS (HP09).
The ratio of mass in the head to tail was investigated for a dependence on other cloud properties such as V$_{\rm GSR}$, peak column density, flux, and size.  The only property that showed a clear trend, was an increasing size (and to a lesser extent an increase in total flux) with an increasing amount of mass in the tail.

\begin{figure}[h!]
\begin{center}
\vspace{-0.01in}
\includegraphics[height=2.5in]{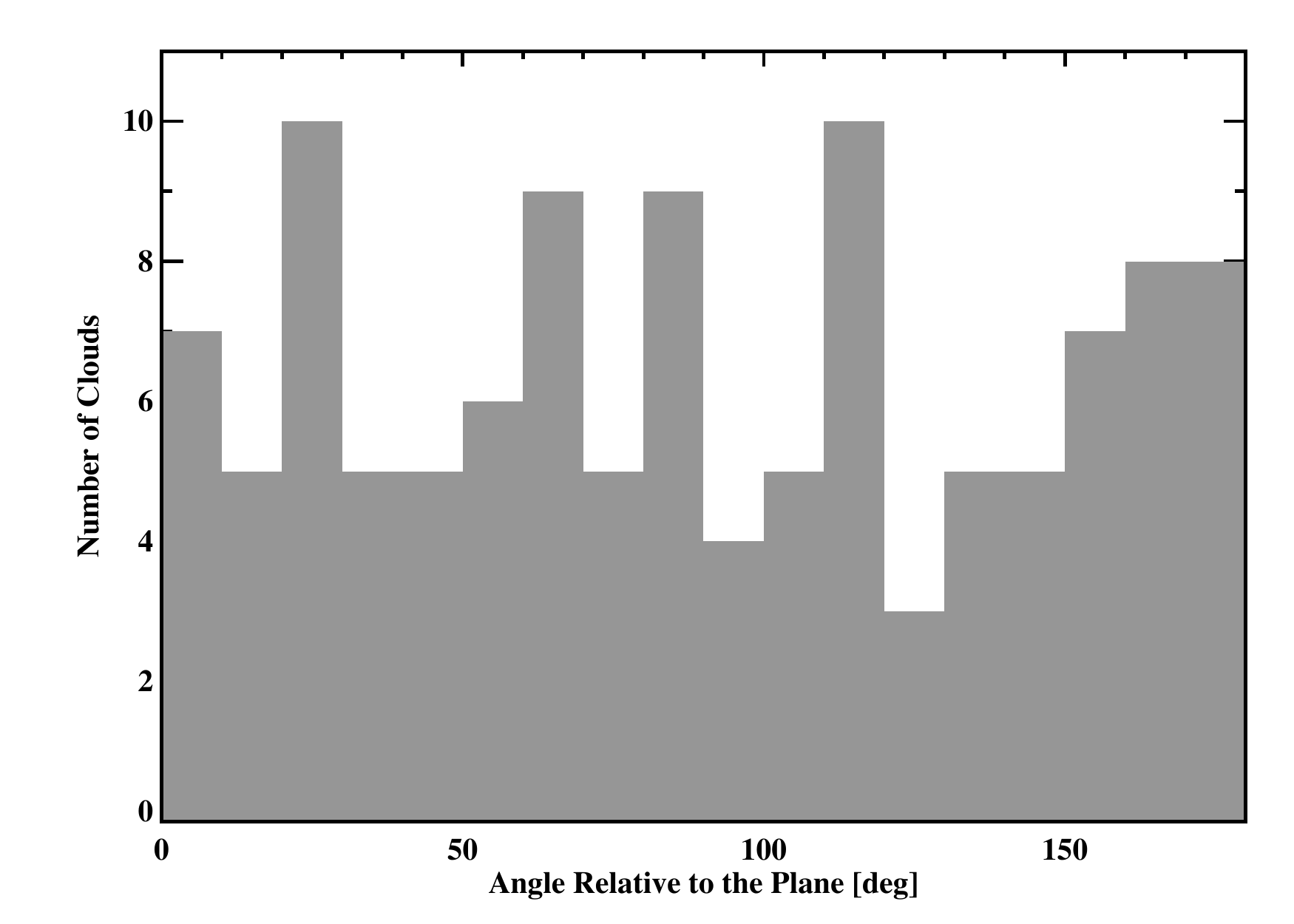}
\end{center}
\caption{The distribution of position angles relative to the Galactic Plane for all head-tail clouds.  90 is parallel to the Plane, 0 infers the head of the cloud is pointed toward the Plane, and 180 is away from the Plane. \label{pa}}
\end{figure}

The distribution of position angles of the head-tail clouds relative to the Galactic Plane is shown in Figure~\ref{pa} with 0\deg~representing clouds with the head pointed towards the plane, 90\deg~parallel to the plane, and 180\deg~clouds that have their head pointed away from the plane.
There is no evidence that the head-tail cloud population as a whole is pointed towards the Galactic plane (or any other direction) when the entire population of head-tail clouds is examined.   This is also
the case when only the negative V$_{GSR}$ clouds are selected.
We also investigated the position angle distribution of the clouds associated with a given complex.  The majority of the head-tail clouds associated with a given complex do not show an obvious position angle alignment, though the small number of clouds associated with many complexes makes this assessment difficult.   We note that for the 12 clouds associated with the Magellanic Stream, 2/3 are pointed generally away from the Plane.  For the clouds associated with the Leading Arm complexes (EP, HP, WC) we investigated the distribution of position angles relative to the North Galactic Pole since these clouds lie on both sides of the Galactic Plane and we wanted to determine if their was an alignment with the motion of the Magellanic System (see \S4.2). 
For these clouds over 3/4 point in the general direction of the NGP rather than the SGP, and 1/2 point within 45\deg~of the NGP.

\section{Discussion}

Head-tail clouds are thought to form as clouds move through a diffuse Galactic halo medium according to the simulations of this process (HP09; \cite{quilis01,konz02, agertz07, santillan01}).
The head of the cloud is compressed due to the pressure at the front of the cloud, and viscous stripping results in a tail of lower column density gas.   We primarily compare our observational results to the simulation results of HP09 throughout this section, as they were set up to simulate exactly the clouds presented here in terms of their observational properties.  These simulations were completed with a three-dimensional grid-based code which includes heating from a UV radiation field and cooling using a metallicity of 10\% solar.   This metallicity is appropriate for some HVCs \citep{collins03}, but may be on the low side for gas originating from the Magellanic System (generally closer to 25\% solar; \cite{gibson00}).    The HP09 simulations include both wind-tunnel and free-fall experiments, or a cloud experiencing a constant density wind and a cloud moving through a 'realistic' halo density profile, respectively.  They were designed to resolve the instabilities acting to disrupt the cloud and cooling within the core of the cloud.  A range of halo densities, cloud masses and velocities were tested with the goal of reproducing the variety of clouds and conditions in the Galactic halo.  The HP09 simulations are the most extensive 3D grid simulations of the disruption of halo clouds currently available (see HP09 for further details).

Much of the variety of structure found in the observed head-tail clouds (e.g., tails with length variation, kinks, and/or clumps) is found in the simulations of clouds moving through a diffuse halo medium.   The kinks and clumps result from the turbulent wake and/or cooling and fragmentation of the gas.  When viewing simulated clouds at various angles, different amounts of the tail are visible, similar to the range of tail lengths in the observations.  Only smooth initial density profiles have been examined in simulations, but the initial structure of the HI in the cloud may also play a role in the appearance of the head-tail clouds.  
The tail lags the head of the cloud as it is stripped behind the moving cloud in the models.   We are only able to examine the velocity structure of 37\% of the head-tail clouds and only measure the component of the velocity along the line of sight.  Given this, we find the tail appears to be lagging for the negative velocity clouds, but not for the majority of the positive velocity clouds.  The positive velocity clouds can be largely associated with the Magellanic System and are most likely dominated by tangential velocities given the Large Magellanic Cloud is moving at 380 \kms~\citep{kallivayalil06}.  However, one should also consider that the head-tail structure could be produced via other mechanisms, such as a filament for which only the high column density end is detected.

The origin and fate of the head-tail clouds is directly related to how long they are expected to last, as this will dictate if they remain in close proximity to their originator and whether they will fuel the disk of the Galaxy directly or be integrated into the multi-phase halo.  In the simulations, the clouds are found to disrupt on relatively rapid timescales due to the dynamical instabilities, independent of including cooling (HP09; \citet{quilis01}).   
The cooling causes fragmentation within the cloud and the fragmentation aids the disruption as the cloud moves through the halo (HP09). 
The rate at which a cloud of a given mass disrupts depends primarily on its velocity and the surrounding halo density, and this can be combined into a ram pressure term (P$_{ram}$).  HP09 examined a range of realistic halo densities and cloud velocities and found clouds modeled after those presented here do not have remaining detectable cold HI gas after 35-160 Myrs.   They were also able to derive a relationship for the disruption length ($D_0$), or the distance the cloud can travel before being completely disrupted (or devoid of detectable HI),
\begin{equation}\label{disrupt}
 D_0 = 3.481 {\rm log M} - (-165.2 + 77.16 {\rm log M} - 18.54({\rm log M})^2 + 86.89 {\rm log P}_{ram})^{1/2} ({\rm kpc}),
\end{equation}
where $D_0$ is in kpc, P$_{ram}$ is in K cm$^{-3}$, and M is the HI mass of the cloud in \Msun.\footnote{This equation is not valid for particularly massive clouds, or low values of P$_{\rm ram}$.  For instance, if a cloud has a mass of $10^5$ \Msun~we can only comment on the disruption length (or time) if P$_{\rm ram} > 10^{2.5}$ K cm$^{-3}$, which is equivalent to the cloud moving at 100 \kms~through a hot halo ($10^6$ K) with a density of 10$^{-3.5}$ cm$^{-3}$ (see figure 7 in HP09).}  
Since the HI mass of a head-tail cloud depends on its distance squared, we discuss the head-tail clouds below in terms of a lower halo ($\sim10$ kpc) and Magellanic population ($\sim60$ kpc) based on their association with HVC complexes (see \S3 and Figures~\ref{complb} and \ref{complv}).  The proximity of the clouds to HVC complexes suggests they originated from a larger HI structure and is consistent with the short lifetimes found in the simulations.   The head-tail clouds not showing distinct properties from the non-head-tail clouds (see Figure~\ref{props} and \S3) may also be consistent with short lifetimes in the sense that once a cloud develops a head-tail structure it rapidly disperses.   The origin and fate of the clouds are discussed further below.

\subsection{Head-Tail Clouds in the Lower Halo}

In this section, we discuss the $\sim$50\% (62/116) of the clouds that are likely to be at distances of $\sim10$ kpc given the distance constraints on some of the associated complexes in Figures~\ref{complb} and \ref{complv} and other HVC complexes \citep{thom06,thom08,wakker01,wakker08,putman03b,weiner01}. 
At 10 kpc the head-tail clouds have a median mass of only 730~\Msun~and halo density estimates at this distance are generally $>10^{-4}$ cm$^{-3}$ \citep{hsu11,gaensler08,mccammon02}.   Therefore, even at the top range of head-tail cloud masses at 10 kpc ($\sim2500$ \Msun),  the clouds will not travel more than a few kpc (Eqtn.~\ref{disrupt}) or last more than 50 Myr (HP09).  This is consistent with the head-tail clouds still being in close proximity to the larger HI complexes from which they originated.  Based on our association criteria (Eqtn.\ref{assoc}), a head-tail cloud must have $\delta v < 35$\kms~if it is a few kpc from an associated HVC complex at 10 kpc.

The total mass of this half of the head-tail cloud population at 10 kpc is $6\times10^4$ \Msun.
Given a 50 Myr disruption time for these head-tail clouds, $\sim10^{6}$ \Msun~of clouds per Gyr need to be generated to have a continuous population.  Several of the larger HVC complexes have HI masses of this order and spread over the Gyr timescale this mass in clouds would not be difficult to generate from gas stripped from satellites and/or possibly cold flow material \citep{grcevich09, keres09}.
The clouds may also be generated through condensations from overdense parts of the multi-phase halo \citep{joung11}, although these overdensities should still be associated with existing complexes.

Given the short lifetimes, the head-tail clouds are unlikely to make it to the Galactic disk as cold star formation fuel.  The position angles not pointing towards the Galactic Plane (Figure~\ref{pa}) may also represent the short lifetimes of the HI clouds, as they should show evidence for infall towards the center of the potential if they lasted long enough.  The gas from the head-tail clouds will either arrive as warm blobs or be integrated into the hot Galactic halo. In the simulations of HP09 that included a realistic halo density, it was found that warm fragments left from the disrupted clouds may begin to recool as they slow and are compressed by a higher density surrounding medium closer to the disk.
There is a large population of cold disk-halo clouds that could potentially represent this phenomenon (Saul et al. 2011; \cite{ford10, lockman02b}).

\subsection{Head-Tail Clouds associated with the Magellanic System}

In the southern hemisphere the dominant high velocity HI features are those stripped from the Magellanic Clouds:  the trailing Magellanic Stream, the Magellanic Bridge joining the LMC and SMC, and the collection of clouds that make up the Leading Arm (see \cite{putmanmagrev} for a review).   In \S3 we identify head-tail clouds as associated with the Magellanic System if they are found within D $< 25$\deg~of the Magellanic Stream (MS) or with complexes EP, HP, and WC in the leading region of the Magellanic System (see Figures~\ref{complb} and \ref{complv}).   This is a total of 54 clouds or close to 50\% of the head-tail clouds that can be associated with the Magellanic System.   The distances to the Magellanic features remain somewhat uncertain.  Since the Stream and the main part of the Leading Arm are connected to the Magellanic Clouds, they must be at $\sim60$ kpc (the distance to the SMC \citep{hilditch05}) near this connection.   The Magellanic Stream extends 100\deg~from the Magellanic Clouds and, according to tidal models of its formation \citep{besla10, gardiner99, connors06b}, may reach distances of 100-120 kpc at its tip.   The Leading Arm feature remains at a distance closer to the Magellanic Clouds in these models.  The majority of the associated head-tail clouds are in the Leading Arm region (42/54) and therefore 60 kpc is used as a canonical distance for the associated head-tail clouds.  We note that the 12 clouds associated with the Magellanic Stream may be more distant.

The majority of the head-tail clouds will have HI masses between
$10^{4-5}$ \Msun~at the 60 kpc distance.  There are no direct measurements of the halo density at the distance of the Magellanic Clouds, but values between $10^{-4}$ and $10^{-5}$ cm$^{-3}$ are consistent with the current observational constraints (Hsu et al. 2011; Sembach et al. 2003\nocite{sembach03,hsu11}).  We can use Equation~\ref{disrupt} to estimate how far the clouds will travel before being completely disrupted.  Since the Magellanic Clouds are traveling at 380~\kms~\citep{kallivayalil06}, values of P$_{ram} > 10^{2.5}$ K cm$^{-3}$ are likely and the clouds will travel less than 12 kpc and last less than 150 Myr.  The head-tail clouds associated with the Magellanic System will therefore not make it to the Galactic disk to fuel star formation, but be integrated into the diffuse hot halo, or, as mentioned in \S4.1, sink to the disk as warm clouds and possibly re-cool there (HP09; \cite{blandhawthorn07}).  One caveat to this is the detection of a coherent magnetic field in a Leading Arm cloud that may provide some stabilization against rapid destruction (\cite{mcclure10}; but see also \cite{stone07}).

It could be we are observing the Magellanic System at a unique time given recent models find the Clouds may be approaching the Galaxy for the first time \citep{besla10}, but if the head-tail clouds are stripped from the system at a relatively constant rate and destroyed in $\sim150$ Myr, this population alone will have contributed at least $10^{7}$ \Msun~of gas to the halo per Gyr.   The abundant head-tail clouds in the Leading Arm region therefore indicate this feature may have been more massive in the past.  The Leading Arm currently has only $3.5 \times 10^7$ \Msun~in HI, while the Magellanic Stream has an HI mass of $2 \times 10^8$ \Msun~(both at a constant distance of 60 kpc), and this mass difference has been posed as somewhat of a puzzle given tidal forces are generally expected to produce symmetric streams.  The low mass of the Leading Arm relative to the Magellanic Stream may be partially explained by the abundant head-tail clouds representing its ongoing destruction as it moves through the Galactic halo.  Indeed, since the Leading Arm is usually at closer radii to the Galactic disk than the Magellanic Stream in the simulations of the interaction of the Magellanic Clouds with the Milky Way \citep{gardiner99,connors06b}, it is likely to be moving through a denser halo medium and should be undergoing more disruption.  

The majority (75\%) of the position angles of the head-tail clouds in the leading region are pointed in the general direction of the North Galactic Pole.  This is consistent with the head-tail clouds pointing in the general direction of the motion of the Magellanic System.   There is less agreement for the head-tail clouds associated with the Magellanic Stream and these clouds may even appear to be pointing back towards the Stream.  
The difference between the Stream and Leading Arm may be due to the Stream showing evidence for being embedded within an extended sheath of low HI column density, largely ionized material \citep{gibson00, braun04,lockman02}.  The clouds associated with the Stream may therefore be somewhat shielded from the hot halo medium.   The Leading Arm does not show evidence for a warm ionized component like the Magellanic Stream does in H$\alpha$ emission \citep{putman03b}, and this is potentially consistent with the HI features leading the Magellanic System not being shielded by a more extended component.  

The head-tail clouds associated with the Magellanic System are additional evidence that the Galactic diffuse halo medium extends out to at least the distance of the Magellanic Clouds.
The origin of this extended halo medium may be a combination of the initial baryon collapse, leftover merger material, and Galactic winds.  High resolution fountain models show that hot gas generally reaches the limits of the simulation box \citep{joung06, deavillez04}, and cosmological simulations with galactic winds find they fill the virial radius and partially escape (e.g., \cite{oppenheimer10}).  Based on the OVI absorption line results for the Milky Way halo, the hot gas covers over 67\% of the sky \citep{sembach03}.  This is consistent with our finding of head-tail clouds across the entire southern sky.
The presence of head-tail clouds around M31 \citep{westmeier05} shows that this halo medium may be a common feature of spiral galaxies.

\section{Summary}

This paper presents a population of 116 head-tail halo clouds that are naturally created with the passage of HI clouds through a diffuse halo medium.   It is found that 35\% of the small, relatively isolated clouds from the P02 HIPASS HVC catalog (the CHVCs and :HVCs) show a spatial head-tail structure at 15.5\arcmin\ resolution.   
The majority of the head-tail clouds can be associated with known HVC complexes including a large fraction ($\sim$50\%) with the Magellanic System.   The head-tail clouds typically have the majority of their mass in the head of the cloud, and their velocity, column density, total flux and size properties are similar to the clouds that were not classified as head-tail clouds.   These results are consistent with the clouds not lasting long once in the head-tail phase; they are still in close proximity to the complexes from which they originated and still show similarities to the clouds not obviously undergoing disruption.  

The head-tail clouds are unlikely to make it to the disk of our Galaxy in cold form based on simulations of these types of clouds passing through a diffuse halo medium (e.g., HP09).  This indicates the clouds are more likely to feed the multi-phase diffuse halo than the Galactic disk directly.  The clouds need to be continually generated unless we are observing the Galactic halo at a special time.  The population that is not clearly associated with the Magellanic System ($\sim60$ clouds) would only require approximately $10^6$ \Msun/Gyr to have a continuous population and should be obtainable from the shearing off from larger complexes that may have originated from satellites or cold flows.  The head-tail clouds associated with the Magellanic System are more massive with their larger distance and are contributing on the order of $10^7$  \Msun/Gyr.  This does not include the mass stripped into the halo directly from the edges of the larger clouds.

Of the 54 head-tail clouds that can be associated with the Magellanic System, the majority (42) are in the region leading the Magellanic Clouds.  The abundance in this region is consistent with the Leading Arm being closer to the Galactic disk and moving through a denser halo medium than the trailing Magellanic Stream.   The Leading Arm may therefore have been more massive in the past, and the existence of the head-tail clouds may help to explain the large mass difference between this feature and the Magellanic Stream.   The leading head-tail clouds also show a bias in their position angles consistent with them following the overall motion of the Magellanic System.  The lack of abundant head-tail clouds associated with the Magellanic Stream may be expected given the Stream shows extended diffuse gas surrounding it \citep{gibson00, braun04}.
The existence of head-tail clouds associated with the Magellanic System is further evidence that the Galaxy's halo medium extends out to at least the distance of the Magellanic Clouds ($\sim60$ kpc).

Future observational plans are to investigate data from the Galactic Arecibo L-band Feed Array HI (GALFA-HI) Survey for head-tail clouds and additional disruption features as the clouds move through the halo (e.g., \cite{peek11,peek07,stanimirovic08}).  The GALFA-HI Survey has a resolution of 4\arcmin~and 0.2 \kms, allowing for an investigation of the detailed spatial and kinematic structure of the clouds and further comparisons to high resolution simulations.   
Since the tail can be heavily suppressed if it is moving towards the observer, these data may reveal a kinematic signature of head-tail clouds, with a lower intensity tail of emission present in the HI profile (HP09).   They may also reveal head-tail structures in the small clouds at the disk-halo interface that could represent the recooled warm fragments of HVCs beginning to fall towards the disk again.
The future with ASKAP will provide additional opportunities to probe the HVCs of the southern sky at a similar resolution and sensitivity to GALFA-HI.
The simulations of halo clouds will also continue to move forward (e.g., Joung, Bryan \& Putman 2011; F. Heitsch, pers. comm.) to further investigate the relation of cold clouds to hot halo gas and the role of clouds in feeding galaxies. 

\acknowledgments
We thank Tobias Westmeier for the HVC map with the LAB data, the HIPASS team, and
Fabian Heitsch, Ryan Joung, and Josh Peek for numerous useful discussions.   We acknowledge funding from
the Research Corporation as part of a Cottrell Scholarship, the Luce Foundation, and NSF CAREER grant AST-0904059.

\bibliography{ref}
\bibliographystyle{apj}

\end{document}